\documentclass[aps,prd,reprint,nofootinbib,longbibliography]{revtex4-1}

\usepackage{amsmath}
\usepackage{mathtools}
\usepackage{graphicx}
\usepackage{verbatim}
\usepackage{ulem}
\usepackage[colorlinks=true, allcolors=blue]{hyperref}
\DeclareUnicodeCharacter{2009}{ }

\begin{document}

\title{Particle-in-cell Simulation of the Neutrino Fast Flavor Instability}
\author{Sherwood Richers}
\email{srichers@berkeley.edu}
\affiliation{Department of Physics, University of California Berkeley, CA 94720}
\affiliation{Nuclear Science Division, Lawrence Berkeley National Lab, CA 94720}
\author{Don E. Willcox}
\affiliation{Computational Research Division, Lawrence Berkeley National Lab, CA 94720}
\author{Nicole M. Ford}
\affiliation{Computational Research Division, Lawrence Berkeley National Lab, CA 94720}
\author{Andrew Myers}
\affiliation{Computational Research Division, Lawrence Berkeley National Lab, CA 94720}
\begin{abstract}
Neutrinos drive core-collapse supernovae, launch outflows from neutron star merger accretion disks, and set the ratio of protons to neutrons in ejecta from both systems that generate heavy elements in the universe. Neutrinos of different flavors interact with matter differently, and much recent work has suggested that fast flavor instabilities are likely ubiquitous in both systems, but the final flavor content after the instability saturates has not been well understood. In this work we present particle-in-cell calculations which follow the evolution of all flavors of neutrinos and antineutrinos through saturation and kinematic decoherence. We conduct one-dimensional three-flavor simulations of neutrino quantum kinetics to demonstrate the outcome of this instability in a few example cases. We demonstrate the growth of both axially symmetric and asymmetric modes whose wavelength and growth rate match predictions from linear stability analysis. Finally, we vary the number density, flux magnitude, and flux direction of the neutrinos and antineutrinos and demonstrate that these factors modify both the growth rate and post-saturation neutrino flavor abundances. Weak electron lepton number (ELN) crossings in these simulations produce both slow growth of the instability and little difference between the flavor abundances in the initial and final states. In all of these calculations the same number of neutrinos and antineutrinos change flavor, making the least abundant between them the limiting factor for post-saturation flavor change. Many more simulations and multi-dimensional simulations are needed to fully probe the parameter space of the initial conditions.
\end{abstract}

\maketitle

\section{Introduction}
Supernovae represent the explosive birth and neutron star mergers the cataclysmic destruction of neutron stars (see \cite{metzger_kilonovae_2017,sarin_evolution_2020,barnes_physics_2020,radice_dynamics_2020,pejcha_explosion_2020,abdikamalov_gravitational_2020,muller_hydrodynamics_2020} for recent reviews). Neutrinos are the dominant couriers of energy and lepton number in both cases and will be observable from the next galactic event. They carry the energy that explodes the star in a core-collapse supernova, drive outflows, and modify the ratio of neutrons to protons available for nucleosynthesis, thereby determining the abundances of elements that enrich the nearby universe. These heavy elements mix with ambient hydrogen that forms later generations of stars and planets.

Although state of the art models in many cases yield qualitatively correct explosion energies, ejecta masses, neutrino signals, electromagnetic transients, and gravitational waves (e.g., \cite{burrows_overarching_2020,oconnor_global_2018,mezzacappa_gravitational-wave_2020,muller_3d_2020,glas_three-dimensional_2019}), there remain many holes in the details. Simulations are growing increasingly sophisticated, but still suffer from low resolution (e.g., \cite{radice_neutrino-driven_2016,nagakura_towards_2019}) and uncertain initial conditions (e.g., \cite{schneider_equation_2020,andresen_gravitational_2019,muller_supernova_2017}). Simulations also require as input an equation of state for matter beyond nuclear densities (e.g.,  \cite{radice_gw170817_2018,margalit_multi-messenger_2019,schneider_equation_2020,richers_equation_2017}) and the nuclear reaction rates of heavy and very unstable elements (e.g., \cite{mumpower_impact_2015}), both of which are poorly constrained. The final dominant model uncertainty is the treatment of neutrinos. The weak interactions between neutrinos and matter and other neutrinos cause them to be largely out of equilibrium, necessitating an expensive kinetic treatment. There is a natural trade-off between inexpensive approximate methods that allow for a larger number of simulations or a focus on specific aspects of the problem and high-accuracy methods that result in different explosion dynamics, ejecta properties, and neutrino signals \cite{richers_detailed_2017,liebendorfer_supernova_2005,oconnor_global_2018,just_core-collapse_2018,glas_three-dimensional_2019,richers_monte_2015,foucart_evaluating_2018,andresen_gravitational-wave_2020}.

In addition, no global simulation dynamically treats the full effects of neutrino flavor transformation. Individual neutrinos exist in a quantum superposition of flavor states, so neutrino masses and potentials due to interactions with matter and other neutrinos can drive rapid and large transformations between quantum states. The neutrino-neutrino interaction term makes the evolution equations nonlinear, resulting in an exponentially difficult many body problem \cite{rrapaj_exact_2020}. Because of this complexity the majority of the work in the field has been done under the mean field approximation (e.g., \cite{vlasenko_neutrino_2014,volpe_neutrino_2015}), though the history and phenomenology is quite rich even with this approximation.

Vacuum and matter-induced oscillations were applied to core-collapse supernovae shortly after they were proposed as a solution to the solar neutrino problem \cite{mikheyev_resonant_1989,wolfenstein_neutrino_1979,wolfenstein_neutrino_1978} and are still used to map supernova simulation output spectra to estimate signals detectable at earth (e.g., \cite{horiuchi_what_2018}). So-called collective oscillations occur when the potential neutrinos feel from interactions with other neutrinos is comparable to the neutrino mass energy scale and has a long history in the core-collapse supernova context (see \cite{duan_collective_2010,bellini_neutrino_2014} for recent reviews). State of the art bulb-model simulations indicate that this transformation mode is likely unimportant for the explosion mechanism (e.g., \cite{duan_self-induced_2011}), but is still impactful for ejecta nucleosynthesis (e.g., \cite{george_fast_2020,grohs_consequences_2020,xiong_potential_2020}). The matter-neutrino resonance can occur under certain conditions when the neutrino potential is comparable to the matter potential and may impact neutrino signals and nucleosynthesis from both supernovae \cite{vlasenko_matter-neutrino_2018} and neutron star mergers \cite{malkus_symmetric_2016}. In addition, a small fraction of neutrinos emitted from an ongoing core-collapse supernova explosion will scatter off of heavy nuclei in the collapsing stellar envelope far outside of the shock front, forming a diffuse halo of scattered neutrinos, some of which are moving inward. These diffuse neutrinos can drive significant flavor changes in the much larger number of outgoing neutrinos \cite{cherry_halo_2013,cherry_time_2020,morinaga_fast_2020}.

Although proposed more than a decade ago \cite{sawyer_speed-up_2005}, it was only appreciated in recent years that a new flavor instability could drive neutrino flavor transformation orders of magnitude more quickly than the previous transformation mechanisms \cite{sawyer_neutrino_2016, chakraborty_self-induced_2016,izaguirre_fast_2017,capozzi_collisional_2019,dasgupta_fast_2018,capozzi_fast_2019,yi_dispersion_2019}. While the magnitude of the effect of this fast flavor instability (FFI) on the abundances of each neutrino flavor is still uncertain, straightforward arguments suggest that the FFI should be a ubiquitous feature of both CCSNe and NSMs \cite{wu_fast_2017,morinaga_fast_2020}. Although the associated small length and timescales make simulations of the full quantum kinetic equations (QKE) presently impossible in global multidimensional simulations, neutrino transport without flavor transformation is much more tractable. Many authors have used the neutrino distributions in these multidimensional neutrino transport simulations to diagnose the presence or lack of the FFI in CCSNe \cite{tamborra_flavor-dependent_2017,abbar_occurrence_2019,m_d_azari_linear_2019,m_delfan_azari_fast_2020,abbar_fast_2020,nagakura_fast-pairwise_2019,xiong_potential_2020,morinaga_fast_2020,glas_fast_2020,capozzi_fast_2020,abbar_characteristics_2020} and NSMs \cite{wu_fast_2017,wu_imprints_2017,george_fast_2020}.  Even if flavor transformation does not affect the dynamics, it has been suggested that it can significantly modify the elements formed in the ejecta, hampering the production of heavy elements in NSMs \cite{wu_imprints_2017,george_fast_2020} and enhancing the production of light-p nuclei in neutrino-driven winds from CCSNe \cite{xiong_potential_2020}.

Motivated by the potential impact of the FFI and the hope that a general understanding of the final state of an unstable distribution can eventually be applied to global simulations, several authors have begun working on direct simulations of the instability. The majority of the simulations so far assume homogeneity (``one-zone'' models) and are discretized in only angle (or angular moments) and time. These simulations have demonstrated a consistency between linear stability analysis and direct evolution of the nonlinear equations \cite{dasgupta_fast_2017,abbar_fast_2019,shalgar_dispelling_2020,johns_fast_2020} and have shed some insight into the late-time angular turbulence and kinematic decoherence \cite{johns_neutrino_2020, bhattacharyya_fast_2020, bhattacharyya_late-time_2020}. Other simulations of the FFI have included inhomogeneity \cite{martin_dynamic_2020,padilla-gay_multi-dimensional_2020}, a simplified treatment of collisional processes \cite{shalgar_change_2020}, or both \cite{capozzi_collisional_2019}.

The capabilities of existing simulation methods are currently limited by various imposed symmetries, a reduced number of neutrino species, and/or a requirement for a very large number of grid cells or basis elements. To unify and expand on these models, a more general framework for simulating kinetics is needed. Neutrino transport methods such as moment methods \cite{chu_thornado-transport_2019,roberts_general-relativistic_2016,oconnor_open-source_2015,oconnor_two-dimensional_2018,vaytet_numerical_2011,foucart_impact_2016,sekiguchi_dynamical_2015,just_new_2015}, discrete ordinates \cite{nagakura_three-dimensional_2019,liebendorfer_finite_2004}, and Monte Carlo \cite{richers_rank-3_2020,foucart_monte-carlo_2020,miller_full_2019,kato_neutrino_2020} could in principle all be extended to treat coherent flavor effects \cite{zhang_transport_2013,richers_neutrino_2019}, but the strong dependence of the evolution of each neutrino on integrals of nearby neutrino distributions is challenging to implement efficiently.

In this paper, we utilize technology from the plasma physics community and describe a particle-in-cell (PIC) method that formally solves the mean field quantum kinetic equations in an efficient, scalable manner. In Section~\ref{sec:methods} we outline a particle-in-cell implementation of the neutrino quantum kinetic equations. In Section~\ref{sec:results} we demonstrate the exponential growth, saturation, and kinematic decoherence of a toy neutrino distribution on a one-dimensional mesh. Finally, in Section~\ref{sec:pstudy} we vary the neutrino distributions to begin the parameter study needed to build a sub-grid model of the FFI. We provide some concluding remarks in Section~\ref{sec:conclusion}.

We have developed the new code {\tt Emu} \cite{willcox_amrex-astroemu_2021} to implement this PIC method for solving the QKEs. {\tt Emu} is fully open-source and is available at \url{https://github.com/AMReX-Astro/Emu}. All parameter files and select data from this study are publicly available \cite{richers_particle--cell_2021} and further data is available upon request.

\section{{\tt Emu}: PIC Neutrino Flavor Kinetics}
\label{sec:methods}
{\tt Emu} solves the mean-field quantum kinetic equations without collisions (Section~\ref{sec:qke}) by evolving the position and quantum state of a collection of computational particles moving through a background grid (Section~\ref{sec:pic_algorithm}). During each simulation timestep the particles aggregate their quantum states to construct a distribution within each grid cell (Section~\ref{sec:pic_deposition}). The neutrino and background matter distribution are next interpolated to each particle's position in order to construct time derivatives of the position and quantum state (Section~\ref{sec:pic_interpolation}). All particles are then integrated forward in time using a high-order integrator and a performance portable domain decomposed parallelization scheme using the {\tt AMReX} framework (Section~\ref{sec:pic_implementation}). This is repeated until the simulation is evolved for the desired amount of time. We discuss each of these steps in more detail below.

\subsection{Quantum Kinetic Equations}
\label{sec:qke}
The quantum kinetic equations that describe the transport of relativistic quantum particles read \cite{vlasenko_neutrino_2014,volpe_neutrino_2015}
\begin{equation}
\label{eq:qke}
    \frac{\partial f_{ab}}{\partial t} + c\mathbf{\Omega} \cdot \mathbf{\nabla} f_{ab} = \mathcal{C}_{ab} - \frac{i}{\hbar}\left[\mathcal{H},f\right]_{ab}\,\,.
\end{equation}
$f_{ab}(\mathbf{x},t,\mathbf{p})$ is a $N_F\times N_F$ Hermitian matrix, where $N_F$ is the number of neutrino flavors. Throughout this paper we use the convention that $a,b,c,d$ are flavor indices, and are $ \in \{e, \mu, \tau\}$ for quantities in the flavor basis or $\in \{1,2,3\}$ for quantities in the mass basis. The diagonals of $f_{ab}$ represent the occupation probability for neutrinos of flavor $a$ located at position $\mathbf{x}$ and time $t$, moving with momentum $|\mathbf{p}|$ in direction $\mathbf{\Omega}=\mathbf{p}/|\mathbf{p}|$ at approximately the speed of light $c$. The neutrino energy is determined by $\epsilon^2=\mathbf{p}^2 c^2 + m^2 c^4 \approx \mathbf{p}^2 c^2$ with energy $h\nu$. One can write a similar equation for the antineutrino distribution $\bar{f}_{ab}$ that involves a distinct collision integral $\bar{\mathcal{C}}_{ab}$ and Hamiltonian $\bar{\mathcal{H}}_{ab}$. Throughout this work we neglect the collision terms $\mathcal{C}_{ab}$ and $\bar{\mathcal{C}}_{ab}$, which are weak enough to not significantly affect the distributions on the timescales simulated here. Implicit in this form of the equations is the assumption that there is no spin coherence, there are no right-handed neutrinos or left-handed antineutrinos, spacetime is flat, and the neutrino momentum is a constant of motion.

The Hamiltonian term $\mathcal{H}_{ab}$ is also a $N_F\times N_F$ Hermitian matrix that encodes potential energy in the form of mass and interactions with other particles. It is usually broken down into a sum of the vacuum potential (due to the neutrino mass), the matter potential (due to interactions with non-neutrino particles), and the self-interaction potential (due to interactions with other neutrinos). In the convention we use these can be written as
\begin{equation}
\label{eq:potentials}
    \begin{aligned}
    \mathcal{H}_{\mathrm{vaccum},ab} &= U_{ac} H^{(m)}_{\mathrm{vaccum},{cd}} U_{db}^\dagger \\
    \mathcal{H}_{\mathrm{matter},ab} &= \sqrt{2} G_F (\hbar c)^3 \left[(n_a - \bar{n}_a^*)-\mathbf{\Omega}\cdot(\mathbf{f}_{a} - \bar{\mathbf{f}}_{a}^*)\right]\delta_{ab} \\
    \mathcal{H}_{\mathrm{neutrino},ab} &= \sqrt{2} G_F  (\hbar c)^3\left[(n_{ab}-\bar{n}_{ab}^*) - \mathbf{\Omega}\cdot(\mathbf{f}_{ab} - \bar{\mathbf{f}}_{ab}^*)\right]
    \end{aligned}
\end{equation}
where $H^{(m)}_{\mathrm{vacuum},ab} = \sqrt{\mathbf{p}^2 c^2 + m_a^2 c^4} \delta_{ab}\approx |\mathbf{p}|c + m_a^2 c^4\delta_{ab}/2|\mathbf{p}| $. The first term ($|\mathbf{p}|c$) is the same for all flavors and therefore cancels out in the commutator in Equation~\ref{eq:qke}, so it can be ignored here. The magnitude of the momentum vector is $|\mathbf{p}|\approx h\nu$, so the vacuum Hamiltonian is usually written as $\mathcal{H}_{\mathrm{vacuum},ab}\approx m_a^2 c^4/2h\nu$. The number density and flux are angular integrals over the neutrino distribution:
\begin{equation}
    \begin{aligned}
        n_{ab}&=\frac{1}{c^3}\int d\mathbf{\Omega}\int d\left(\frac{\nu^3}{3}\right) f_{ab}\,\,, \\
        \mathbf{f}_{ab}&=\frac{1}{c^3}\int d\mathbf{\Omega}\int d\left(\frac{\nu^3}{3}\right) f_{ab} \mathbf{\Omega}\,\,.\\
    \end{aligned}
\end{equation}
Note that we denote a scalar lepton number density with $n_a$ and a Hermitian matrix neutrino number density with $n_{ab}$. $\mathbf{f}_{ab}$ is the (vector of Hermitian matrices) neutrino number flux density, not to be confused with the distribution function $f_{ab}$ or the lepton flux $\mathbf{f}_a$. $\bar{n}_a$ and $\bar{\mathbf{f}}_a$ are real, but we leave the complex conjugation on them in Equation~\ref{eq:potentials} for comparison to the neutrino potential. Our simulations are performed in the fluid comoving frame, making $\mathbf{f}_a=\bar{\mathbf{f}}_a=0$. For antineutrinos, $\bar{\mathcal{H}}_{\mathrm{vacuum},ab}=\mathcal{H}_{\mathrm{vacuum},ab}^*$, $\bar{\mathcal{H}}_{\mathrm{matter},ab} = -\mathcal{H}_{\mathrm{matter},ab}^*$, and $\bar{\mathcal{H}}_{\mathrm{neutrino},ab} = -\mathcal{H}_{\mathrm{neutrino},ab}^*$. Note also that there are other conventions in the literature that affect the form of the antineutrino Hamiltonians depending on whether one writes the evolution equation for $\bar{f}_{ab}$ or $\bar{f}_{ab}^*$.

\subsection{PIC Algorithm}
\label{sec:pic_algorithm}
The left hand side of Equation~\ref{eq:qke} is a comoving derivative moving along the neutrino trajectory. Therefore rather than evaluating the derivative as fluxes between adjacent grid cells as is done in a finite volume scheme, we can simulate the distribution as a set of particles such that the advection term is accounted for by moving the particle's location. A computational particle carries two scalars $N$ and $\bar{N}$ that represent the number of physical neutrinos and antineutrinos contained in the computational particle, along with two $N_F\times N_F$ Hermitian unit-trace quantum density matrices $\rho$ and $\bar{\rho}$ that represent the quantum state of each neutrino and antineutrino. In practice, we only store and evolve the real and imaginary components of each element in the upper right half and the diagonal of these Hermitian matrices in the flavor basis. Wherever the lower left corners of Hermitian matrices are required, we express them in terms of the components of the upper right half in our implementation.

For a given particle labeled by index $p$, the equations of motion can then be expressed as
\begin{equation}
\begin{aligned}
\label{eq:eom}
    \frac{\partial \rho_{ab,p}}{\partial t} &= -\frac{i}{\hbar}\left[\mathcal{H}_p,\rho_p\right]_{ab}\\
    \frac{\partial \bar{\rho}_{ab,p}}{\partial t} &= -\frac{i}{\hbar}\left[\bar{\mathcal{H}}_p,\bar{\rho}_p\right]_{ab}\\
    \frac{\partial \mathbf{x}_p}{\partial t} &= c \mathbf{\Omega}_p\\
    \frac{\partial \nu_p}{\partial t} &= 0 \\
    \frac{\partial \mathbf{\Omega}_p}{\partial t} &= 0\\
    \frac{\partial N_p}{\partial t} &=\frac{\partial \bar{N}_p}{\partial t}= 0 \\
\end{aligned}
\end{equation}
Formally, these equations are only true in the limit of a large neutrino energy compared to the neutrino mass and potentials, but the same is true of Equation~\ref{eq:qke}. In both forms, more terms would need to be included to faithfully treat low-energy neutrinos (e.g. \cite{vlasenko_neutrino_2014}).

Because the neutrino self-interaction couples the flavor evolution of neutrinos and antineutrinos to their local distributions, we integrate Equation~\ref{eq:eom} for all particles together in step.
To compute the right hand side of Equation~\ref{eq:eom} for each particle, we need to evaluate the Hamiltonian terms in Equation~\ref{eq:potentials}.  The matter and self-interaction potentials respectively depend on the local lepton and neutrino density and flux. Because evaluating the local neutrino density and flux requires determining the local neutrino distributions represented by the particles, we evaluate the Hamiltonian in two steps: deposition and interpolation.

\subsubsection{Evaluating the Hamiltonian: Depositing Particle Information to the Grid}
\label{sec:pic_deposition}

The background neutrino density stored in each grid cell must reflect the particle distributions themselves. A straightforward way would be to simply add up all of the particles within a cell, which would assume that each particle is shaped as a delta function in position and only interacts with particles in the same grid cell. However, we found that this causes the particle evolution to be unstable to artificially growing perturbations with wavelengths on the length scale of the grid resolution. We eliminated this numerical instability by instead modelling each particle with an extended shape as is usually done in PIC methods in order to smooth particle-mesh deposition and interpolation operations with higher order stencils.

In plasma PIC simulations, computational particles can be designed to represent a distribution of charge centered at the particle position but with an extent comparable to the size of a grid cell that defines the particle's interpolating ``shape'' function. This yields higher order spatial convergence and suppression of spurious numerical instabilities. We apply the same method in {\tt Emu} for particle flavor matrices instead of particle charge. The particle shape routines in {\tt Emu} are copied directly from the open-source plasma physics code {\tt WarpX}\footnote{https://github.com/ECP-WarpX/WarpX} \cite{vay_warp-x_2018} and include options from a delta function (zeroth order interpolant) to a cubic distribution (third order interpolant). We generally use the quadratic shape, since the cubic shape requires more ghost zones (see Section~\ref{sec:pic_implementation}) and higher-order interpolators do not necessarily correspond to higher-order convergence in PIC codes \cite{wang_particle--cell_2011,edwards_high-order_2012}. We integrate the total neutrino density and neutrino number flux at grid cell $j$ as
\begin{equation}
\label{eq:pic_deposition}
\begin{aligned}
    n_{ab,j} &= \frac{1}{\Delta V}\sum_{p=1}^{N_\mathrm{particles}} N_p w_{pj}\rho_{ab,p} \\
    \mathbf{f}_{ab,j} &= \frac{1}{\Delta V}\sum_{p=1}^{N_\mathrm{particles}} N_p w_{pj} \rho_{ab,p} \mathbf{\Omega}_p
    \end{aligned}  
\end{equation}
where $w_{pj}$ is the fraction of particle $p$ that contributes to grid cell $j$ according to the particle's shape function. $N_\mathrm{particles}$ is the total number of computational particles in the simulation, $N_p$ is the number of physical neutrinos that particle $p$ represents, and $\Delta V$ is the volume of the grid cell.

\subsubsection{Evaluating the Hamiltonian: Interpolating from the Grid to Calculate Neutrino Potentials}
\label{sec:pic_interpolation}

Now that the lepton and neutrino number density and number flux are known at the center of each grid cell, we interpolate them from the grid to each particle's position $\mathbf{x}_p$ using the particle shape function as in Equation~\ref{eq:pic_interpolation}:
\begin{equation}
\label{eq:pic_interpolation}
\begin{aligned}
    n_{a,p}  &= \sum_{j=1}^{N_\mathrm{zones}} w_{pj} n_{a,j}\,\,, \\
    n_{ab,p} &= \sum_{j=1}^{N_\mathrm{zones}} w_{pj} n_{ab,j}\,\,, \\
    \mathbf{f}_{ab,p} &= \sum_{j=1}^{N_\mathrm{zones}} w_{pj} \mathbf{f}_{ab,j}\,\,.
    \end{aligned}  
\end{equation}
This yields the lepton density ($n_{a,p}$), the neutrino densities ($n_{ab,p}$ and $\bar{n}_{ab,p}$) and fluxes ($\mathbf{f}_{ab,p}$ and $\mathbf{\bar{f}}_{ab,p}$). We then obtain the neutrino and antineutrino Hamiltonians $\mathcal{H}_{ab,p}$ and $\bar{\mathcal{H}}_{ab,p}$ for each particle by applying the particle state and these interpolated quantities to Equation~\ref{eq:potentials}. We finally use Equation~\ref{eq:eom} to calculate the particle's density matrix time derivatives
$\partial \rho_{ab,p} / \partial t$ and 
$\partial \bar{\rho}_{ab,p} / \partial t$.

\subsection{PIC Implementation}
\label{sec:pic_implementation}

\begin{figure*}
    \centering
    \includegraphics[width=\textwidth]{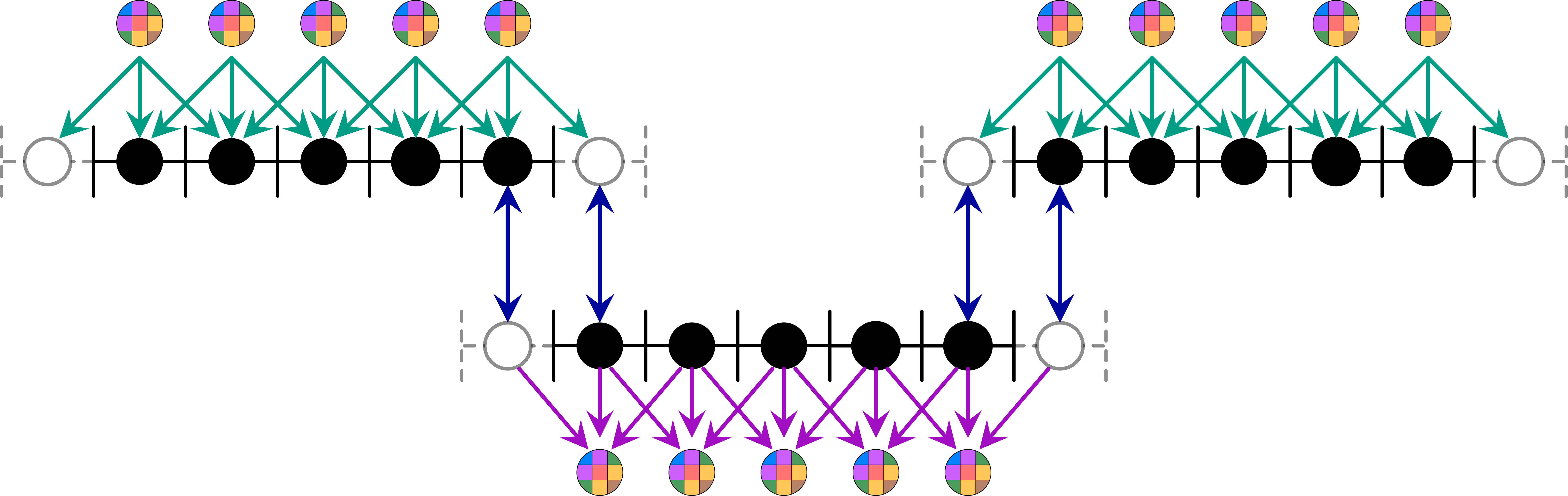}
    \caption{Illustration of the {\tt Emu} algorithm for evaluating the time derivatives of all particles' density matrices. Particles are represented by the colored circles denoting a $3\times 3$ matrix. Each block of grid cells on the domain is represented by a contiguous array of cells composed of interior cells (black circles) and ghost cells (white circles). In the first step of the right hand side evaluation, we deposit particle states onto the underlying grids using second order particle-in-cell shape functions (green arrows, omitted for the center block to avoid cluttering the illustration). This yields neutrino density and fluxes on the grid. We next sum ghost cells into the valid cells on neighboring blocks and then fill ghost cells from the valid cells (blue arrows). Finally, we interpolate neutrino density and fluxes from the grid to particles (purple arrows, omitted for the left and right blocks to avoid cluttering the illustration), construct particle Hamiltonians, and evaluate neutrino density matrix time derivatives. Figure available under CC BY 4.0 license~\cite{noauthor_emu_2021}.}
    \label{fig:emu_eval_rhs}
\end{figure*}

We implement this algorithm in modern C++ using the exascale computing framework AMReX \cite{zhang_amrex_2019}, which provides high level abstractions for domain decomposition and both distributed and shared-memory parallelization. We thus represent the domain as a union of non-overlapping rectangular blocks, each containing particles and Cartesian grid data. These blocks are distributed across MPI ranks, which allocate memory for their local grid and particle data. Because higher order particle shape functions require particles to deposit and interpolate from neighboring cells, when a particle is near a block's boundary it needs to access data stored on other blocks. Another way to say this is that the particle-mesh operations have a spatial stencil width, where the stencil width refers to the number of grid zones in each dimension a particular operation requires. Piecewise constant shape functions have a stencil width of 1, linear shape functions have a stencil width of 2, quadratic shape functions have a stencil width of 3, etc. (centered on the grid zone containing the particle). Rather than communicate data between blocks every time an individual particle requires it, we instead place $N_G$ layers of ghost zones around the boundary of each block to serve as temporary storage. Ghost zones attached to a given block line up with valid zones within the domain of other blocks, so MPI calls can be used to transfer data to and from all ghost zones and their corresponding valid zones all at once. We choose the number of ghost cells consistent with the stencil width of the particle shape function desired (i.e. $N_G=1$ for quadratic shape functions with a stencil size of 3, since the particle-mesh operations require data from the particle's current grid cell and one cell to the left and right in each coordinate direction.).

We illustrate how we implement the deposition, communication, and interpolation steps discussed in Section~\ref{sec:pic_algorithm} in Figure~\ref{fig:emu_eval_rhs}. We carry out the deposition in Equation~\ref{eq:pic_deposition} by first summing (green arrows) particle contributions to neutrino density and flux into local grids, including both interior grid cells (black circles) and neighboring ghost cells (white circles). This happens on all blocks, but we omit green arrows from the center block to avoid cluttering the figure. We next communicate (blue arrows) ghost cell values to the MPI ranks containing the corresponding valid cells and sum the ghost cell contributions into the valid cells. We then copy the grid data from the valid cells to all corresponding ghost cells. We apply periodic boundary conditions to ghost cells at the domain boundaries, mapping them to valid cells on the other side of the domain. Finally, we interpolate (purple arrows) the neutrino density and flux from each grid to particle locations and compute the density matrix right hand sides in Equation~\ref{eq:eom}, as described in Section~\ref{sec:pic_interpolation}. This happens on all blocks, but we omit purple arrows from the left and right blocks to avoid cluttering the figure.

We use a fourth-order explicit Runge-Kutta method to integrate Equation~\ref{eq:eom} for all particles. We choose a timestep based on two criteria as:
\begin{equation}
\label{eq:timestep}
\begin{aligned}
  \Delta t_{\mathrm{advection}} &= C \dfrac{\mathrm{min}(\Delta x, \Delta y, \Delta z)}{c} \\
  \Delta t_{\mathrm{matter}, \nu} &= C_F \dfrac{\hbar}{\sqrt{2} G_F \mathrm{max}(n_{aa}, \bar{n}_{aa}, n_a)} \\
  \Delta t &= \mathrm{min}(\Delta t_{\mathrm{advection}}, \Delta t_{\mathrm{matter}, \nu})\,\,,
  \end{aligned}
\end{equation}
where the maximum is taken over grid cells and matrix components. 
The two constants $C$ and $C_F$ controlling the timestep correspond to the advection and commutator terms in Equation~\ref{eq:qke}, respectively. $C$ and $C_F$ are thus analogous to hydrodynamic CFL factors, and for the simulations in this paper we use $C=C_F=0.5$.
We find that at least a fourth-order Runge-Kutta method is necessary to adequately control the size of the error for the flavor transformation of individual particles. However, {\tt Emu}'s overall order of space-time convergence is formally limited by the second order PIC method as mentioned in Section~\ref{sec:pic_deposition}. We assess {\tt Emu}'s convergence properties for the simulations of Section~\ref{sec:results} in Appendix~\ref{sec:convergence}.

We integrate particle positions concurrently with their flavor states, so we do not split the advection and flavor-changing operators in Equation~\ref{eq:qke}.
Thus, when constructing the state at which to evaluate a Runge-Kutta stage right hand side, we first calculate particle positions along with their density matrices at that time. As a result, a particle may move from one block on the domain to another, which may reside on a different MPI rank. Such a particle will therefore need access to grid data on a different block before we can evaluate Equation~\ref{eq:eom} for this Runge-Kutta stage. Thus, immediately after calculating the new particle positions and density matrices corresponding to a Runge-Kutta stage, if any particle position now resides within a different block we redistribute its data for all Runge-Kutta stages to its new block and MPI rank. We then resume calculating the right hand side defined by Equation~\ref{eq:eom} for this Runge-Kutta stage using the deposition and interpolation operations discussed in Section~\ref{sec:pic_deposition} and Section~\ref{sec:pic_interpolation}. We repeat this redistribution procedure at all Runge-Kutta stages and at the end of the time step. This ensures that we are able to evaluate all Runge-Kutta stage right hand sides for the particles even if previous stage right hand side terms were originally evaluated on different MPI ranks.

Local particle arithmetic operations (deposition, interpolation, evolution) constitute the vast majority of the computational cost. We implement these kernels using AMReX's performance-portable strategy for targeting either CPU threading (with OpenMP) or GPUs (with CUDA). The quantities evolved in Equation~\ref{eq:eom} are arranged in a struct for each particle and each block stores an array of these particle structs. For CPUs, we distribute blocks with particles to threads using OpenMP. Each thread loops over particles within the block, though we find in practice that increasing the number of MPI ranks can give better performance on CPUs than OpenMP threading. For GPUs, we assign one MPI rank to each GPU and allocate local grid and particle memory using CUDA Unified Memory. Each MPI rank loops over its local blocks and asynchronously launches a CUDA kernel parallelizing the loop over particles within the block. We then synchronize GPU kernels only after looping over all local blocks to ensure high kernel occupancy.

\section{Results}
\label{sec:results}

In this work we restrict ourselves to simulations performed in one spatial dimension and two angular dimensions. That is, we arbitrarily assume all neutrinos have an energy of $50\,\mathrm{MeV}$, though this only influences the vacuum Hamiltonian, which has negligible effect on the timescale of these simulations. In this section we examine in detail only a single choice of initial conditions. This simplified setup will serve as a basis for comparison with other simulations in this and future work. For our fiducial simulation we choose initial conditions that exhibit a strong electron lepton number (ELN) crossing (i.e., a direction along which there are equal numbers of electron neutrinos and antineutrinos), characterized by electron neutrino and electron antineutrino distributions with a flux factor of $1/3$ in opposite directions. The distribution is thus quite unstable to the FFI. We randomly perturb the initial conditions to seed all modes, as we will describe in Section~\ref{sec:initialConditions}. The amplitudes of the unstable modes grow exponentially precisely in the way predicted by linear stability analysis as shown in Section~\ref{sec:linearGrowth}. The instability then saturates, diffusing neutrino flavor fluctuations away from the fastest growing wavelength into modes of all wavelengths as described in Section~\ref{sec:saturation}. Select data and all parameter files from this study are publicly available \cite{richers_particle--cell_2021}. Additional data is available upon request.

\subsection{Simulation Setup}
\label{sec:initialConditions}
In our fiducial simulation we use a simulation domain 64 cm in length with a grid resolution of $1\times1\times1024$. The unit size in the $x$ and $y$ directions makes the simulation behave in a one-dimensional manner. That is, the neutrino distributions are homogeneous in the $x$ and $y$ directions, but are not restricted to isotropy or axial symmetry. We place many particles at the center of each grid cell with directions distributed approximately isotropically using the ``Polar Coordinate Subdivision'' method of Ref.~\cite{katanforoush_distributing_2003}, except that we specify the number of points along the equator rather than the total number of points. Requesting 16 directions in the $\hat{x}-\hat{y}$ plane results in 92 particles per grid cell. In contrast, PIC codes often distribute particles initially randomly. We elect to distribute particles regularly so we do not have to consider Monte Carlo numerical noise when interpreting the results.

We arbitrarily choose an initial neutrino distribution with a number density of $n=4.89\times10^{32}\,\mathrm{cm}^{-3}$ of each electron neutrinos and electron antineutrinos, and zero density for heavy lepton neutrinos. The fastest growing mode of a two-beam model with these neutrino densities has a wavelength of $1\,\mathrm{cm}$, allowing a more straightforward comparison to the two-beam test in Appendix~\ref{sec:two_beam_ffi}. To create the desired anisotropic neutrino distribution using an isotropic distribution of computational particles, we give each particle weights (i.e. the number of physical neutrinos or antineutrinos that a computational particle represents) of
\begin{equation}
\begin{aligned}
    N_p = \frac{1}{N_\mathrm{ppz}}(n_{\mathrm{input}} + 3 \mathbf{f}_{\mathrm{input}}\cdot\mathbf{\Omega}) \Delta V\,\,,\\
    \bar{N}_p = \frac{1}{N_\mathrm{ppz}}(\bar{n}_{\mathrm{input}} + 3 \bar{\mathbf{f}}_{\mathrm{input}}\cdot\mathbf{\Omega}) \Delta V\,\,,
    \end{aligned}
\end{equation}
where $n_{\mathrm{input}}$ ($\bar{n}_{\mathrm{input}}$) is the input number density of (anti-)neutrinos and $\mathbf{f}_{\mathrm{input}}$ ($\bar{\mathbf{f}}_{\mathrm{input}}$) is the input number flux of (anti-)neutrinos and $N_\mathrm{ppz}$ is the number of computational particles per grid zone. We use $\mathbf{f}_{\mathrm{input}}=n_{\mathrm{input}} \hat{z}/3$ and $\bar{\mathbf{f}}_{\mathrm{input}}=-\bar{n}_{\mathrm{input}}\hat{z}/3$ for this fiducial simulation. This defines $\hat{z}$ as our axial symmetry axis and results in a healthy ELN crossing in the $\hat{x}-\hat{y}$ plane. The initial state of each particle is set to be in a nearly pure electron flavor state with a random perturbation in the off-diagonal elements. That is,
\begin{equation}
\label{eq:initial_flavor_state}
    \rho_{ab,p} = \begin{bmatrix}
    1-\epsilon_\mu-\epsilon_\tau & \alpha(U+Ui) & \alpha(U+Ui) \\
    \rho_{e\mu}^* & \epsilon_\mu & 0 \\
    \rho_{e\tau}^* & 0 & \epsilon_\tau \\
    \end{bmatrix}\,\,.
\end{equation}
$U \in [-1,1]$ is a uniform random number generated individually each time it appears and $\alpha=10^{-6}$ is the strength of the random perturbation. $\epsilon_\mu$ and $\epsilon_\tau$ are determined after the random numbers are generated in order to ensure unit trace and a flavor vector length of one. Since the initial distributions have almost no $\nu_\tau$ or $\nu_\mu$ content, we set the $\rho_{\mu\tau}$ off-diagonal components to zero.

\begin{figure*}
    \centering
    \includegraphics[width=\textwidth]{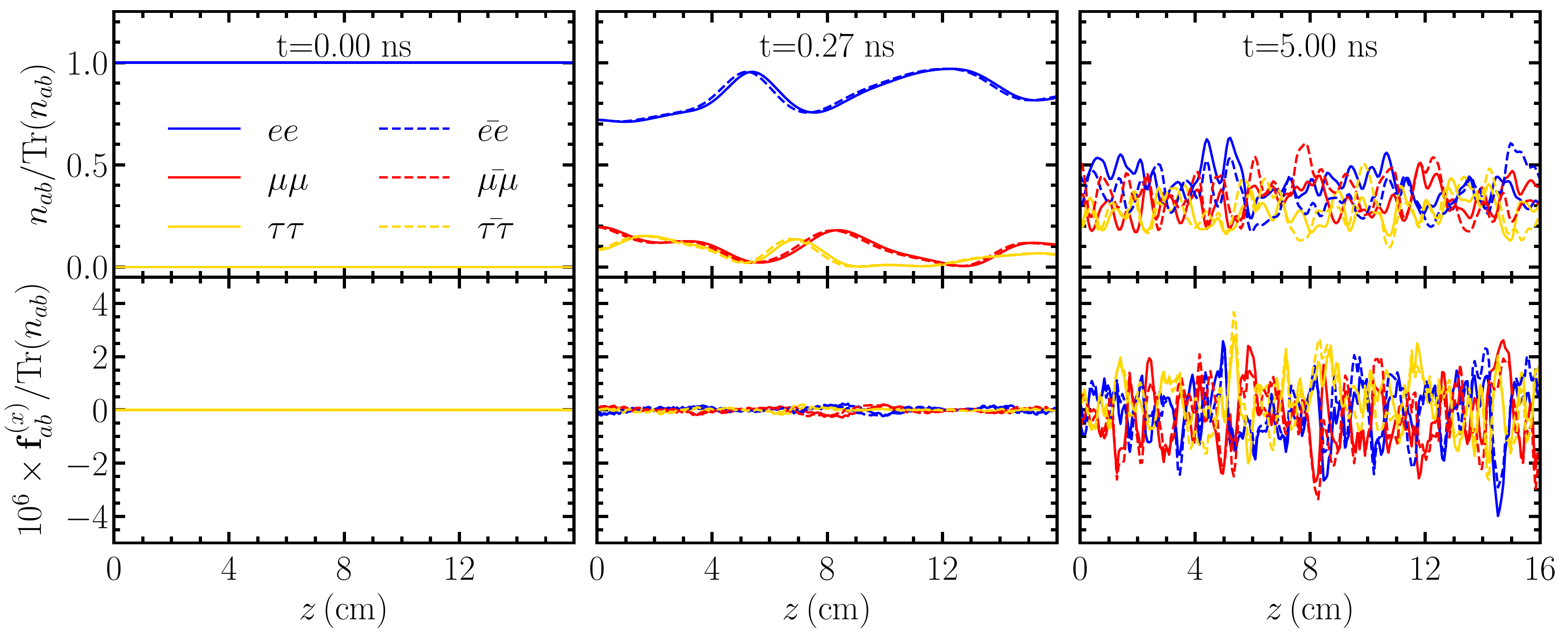}
    \caption{Snapshots of flavor-diagonal components of direction integrated quantities in the fiducial 1D simulation. The three columns show snapshots at the initial time, near the end of the linear growth phase, and the post-saturation phase, respectively. The top row shows neutrino number density and the bottom row shows the flux in the $\hat{x}$ direction. Blue, red, and gold solid (dashed) curves show electron, muon, and tauon flavor (anti-)neutrinos, respectively. The abundances of all flavors are roughly equal in the post-saturation state.}
    \label{fig:1d_diag}
\end{figure*}
\begin{figure*}
    \centering
    \includegraphics[width=\textwidth]{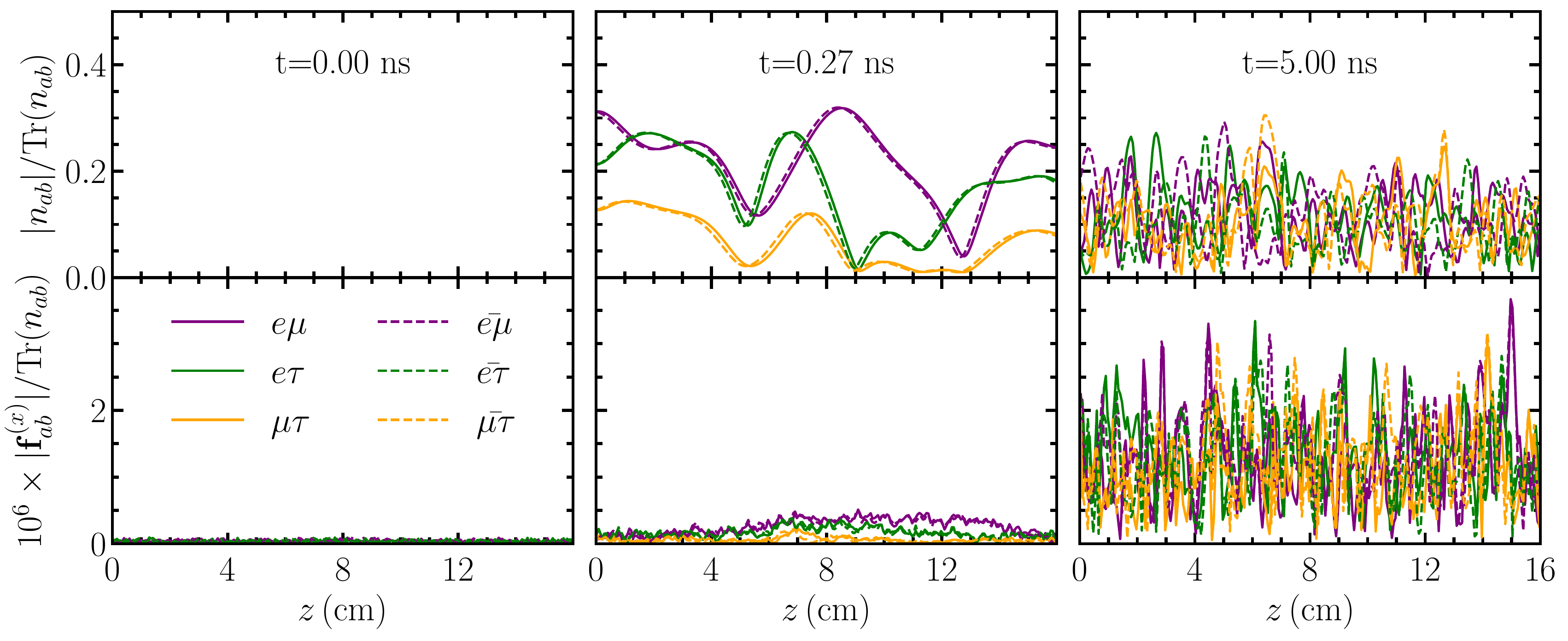}
    \caption{Snapshots of magnitude of flavor off-diagonal components of direction integrated quantities in the fiducial 1D simulation. The three columns show snapshots at the initial time, near the end of the linear growth phase, and the post-saturation phase, respectively. The top row shows neutrino number density and the bottom row shows the flux in the $\hat{x}$ direction. Purple, green, and orange solid (dashed) curves show $e\mu$, $e\tau$, and $\mu\tau$ components for (anti-)neutrinos, respectively. Modes with wavelength near that of the fastest growing mode dominate the linear evolution, but break up into a broad spectrum of wavelengths in the post-saturation phase.}
    \label{fig:1d_offdiag_mag}
\end{figure*}
\begin{figure*}
    \centering
    \includegraphics[width=\textwidth]{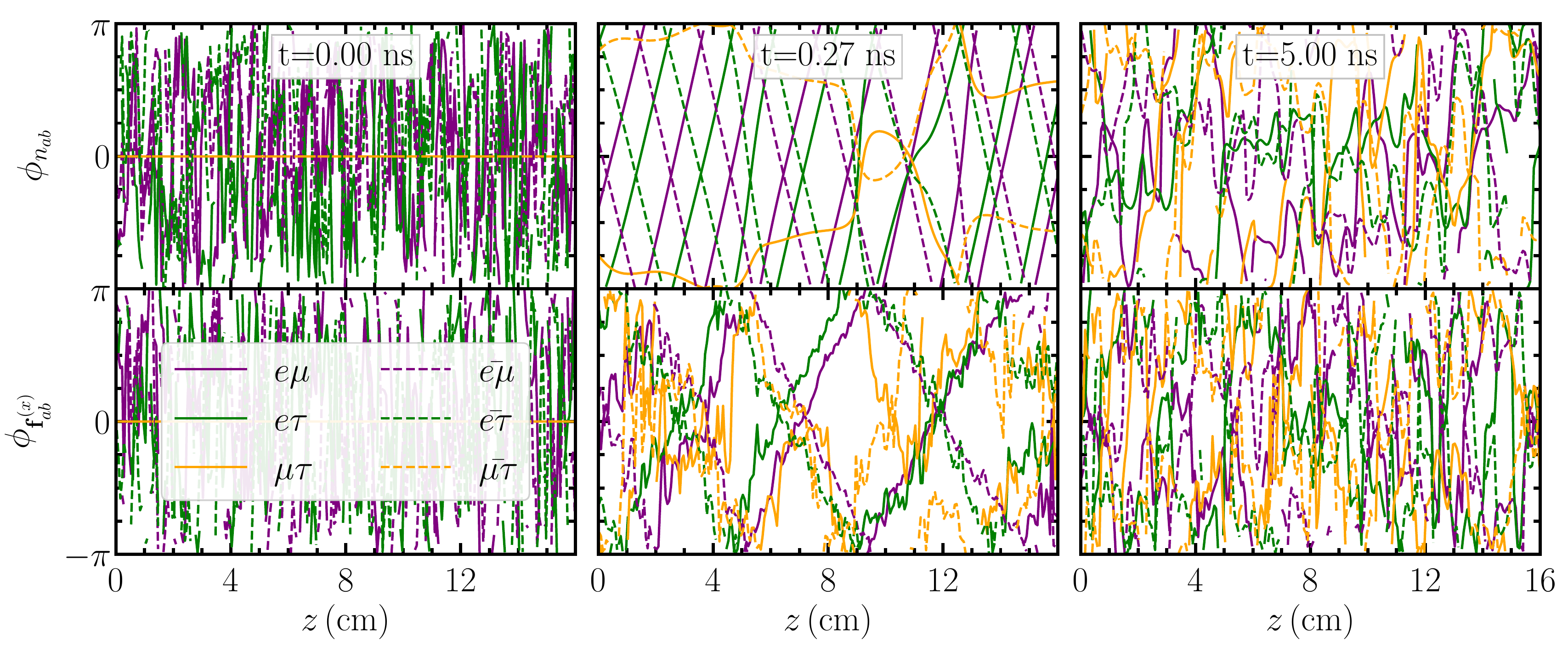}
    \caption{Snapshots of complex phase of flavor off-diagonal components of direction integrated quantities in the fiducial 1D simulation. The three columns show snapshots at the initial time, near the end of the linear growth phase, and the post-saturation phase, respectively. The top row shows neutrino number density and the bottom row shows the flux in the $\hat{x}$ direction. Purple, green, and orange solid (dashed) curves show $e\mu$, $e\tau$, and $\mu\tau$ components for (anti-)neutrinos, respectively. The wavelength (distance between points of equal phase) of the fastest growing axially symmetric mode is visible in the top panel of the center column, while the wavelength of the fastest growing axially asymmetric mode is visible in the bottom panel of the center column.}
    \label{fig:1d_offdiag_phase}
\end{figure*}

The left column of Figures~\ref{fig:1d_diag}-\ref{fig:1d_offdiag_phase} show these initial conditions at $t=0$. The top left panel of Figure~\ref{fig:1d_diag} shows the initial number density of each neutrino species plotted over 16 cm of the 64 cm domain divided by the trace of the number density matrix. All neutrinos (solid curves) and antineutrinos (dashed curves) begin in the electron flavor state (blue), and hence the electron neutrinos and antineutrinos constitute essentially all of the number density. All neutrino flux is in the $\pm \hat{z}$ direction, so the $x$ component of the flux shown in the bottom panel at $t=0$ is zero everywhere. The left column of Figures~\ref{fig:1d_offdiag_mag} and \ref{fig:1d_offdiag_phase} show the magnitude and complex phase, respectively, of the ${e\mu}$ (purple), ${e\tau}$ (green), and ${\mu\tau}$ (orange) components of the neutrino density. The perturbations are too small to be visible in the top left panel of Figure~\ref{fig:1d_offdiag_mag}. Although the perturbations of the density matrix of each particle individually are $\mathcal{O}(10^{-6})$, they average together to make the perturbation of the $x$ component of the flux factor (bottom panel) much smaller than $10^{-6}$. The left panels of Figure~\ref{fig:1d_offdiag_phase} simply show a random distribution of phases of the $e\mu$ and $e\tau$ components of both the number density and flux flavor matrices. The complex phase of the $\mu\tau$ component is shown as zero because it has exactly zero magnitude.

\subsection{Linear Growth Phase}
\label{sec:linearGrowth}

\begin{figure}
    \centering
    \includegraphics[width=\linewidth]{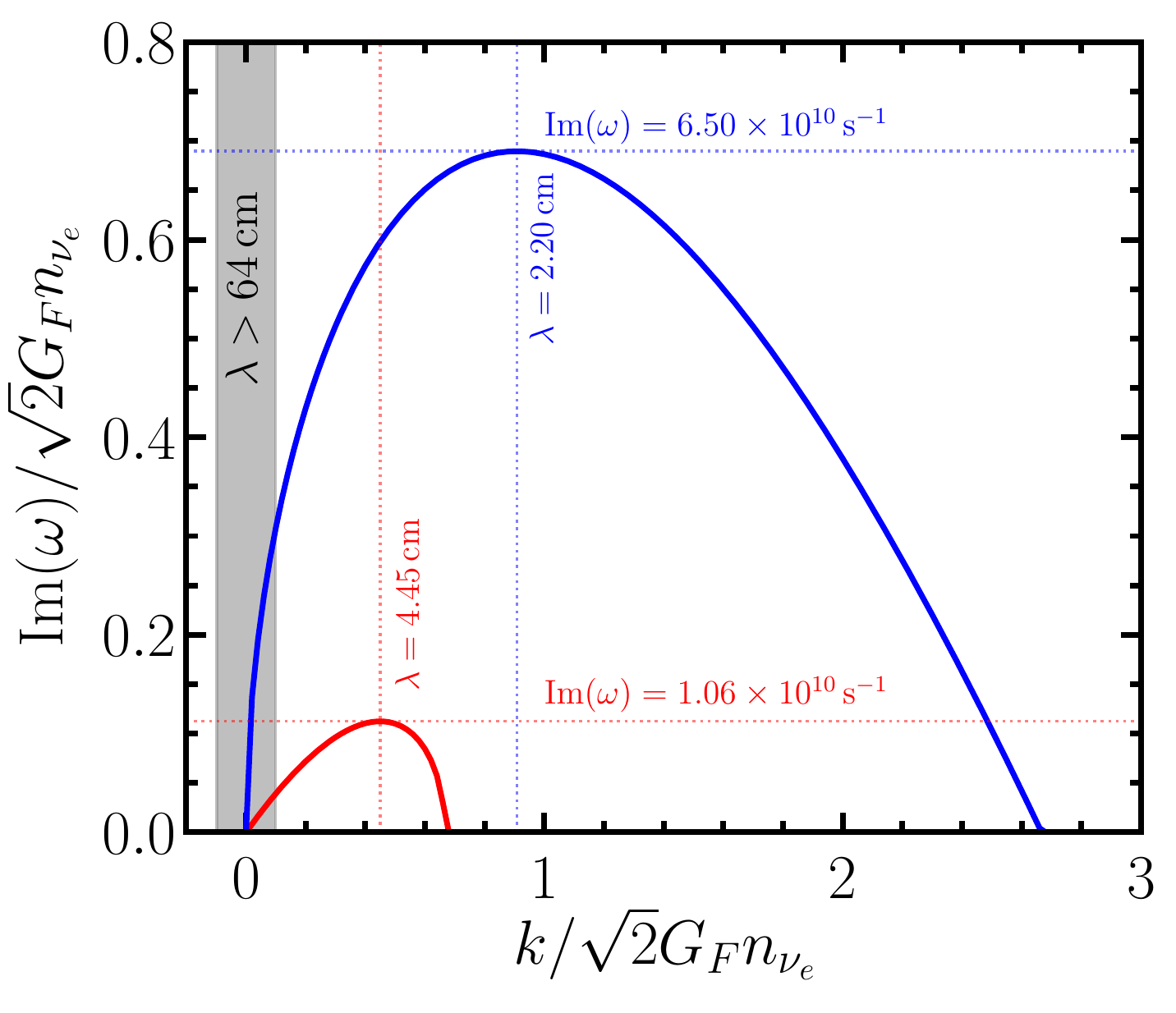}
    \caption{Dispersion relation of modes with frequency $\omega$ and wavenumber $k$. The blue curve corresponds to modes that do not break axial symmetry, while the red curve corresponds to modes that do. Both are normalized by a characteristic self-interaction potential so the diagram is independent of neutrino density as long as the self-interaction potential is much larger than the vacuum potential. Unstable modes exist for wavelengths $\lambda \gtrsim 0.75\,\mathrm{cm}$ ($k/\sqrt{2}G_F n_{\nu_e}\lesssim 2.66$), though the growth rate for the homogeneous mode ($k=0$) is also 0. The gray shadow shows the region of wavenumbers where the corresponding mode cannot exist on our domain because of our limited domain size of $64\,\mathrm{cm}$.}
    \label{fig:dispersion_relation}
\end{figure}
Before discussing the simulation results, we briefly go through the standard linear stability analysis to point out what modes should be unstable and the rate at which they are expected to grow. Following \cite{izaguirre_fast_2017}, we calculate the imaginary component of the frequency for modes of the form $N_{ex}=\alpha e^{-i(\omega t - k z)}$ described by a frequency $\omega$, wavenumber $k$, and initial perturbation amplitude $\alpha$. $x$ here represents either $\mu$ or $\tau$. This is a two-flavor analysis, so the results are applicable to $e-\mu$ and $e-\tau$ flavor transformation but does not describe $\mu-\tau$ transformation \cite{chakraborty_three_2020}. The cosine dependence of the distribution on the polar angle makes computing the integrals in the polarization tensor analytic, though a numerical root find still needs to be performed to extract the growth rate of each mode. We plot the results in Figure~\ref{fig:dispersion_relation}. Since the results scale with the neutrino density (assuming the flux factors and directions are held constant), we plot both the growth rate and wavenumber scaled by $\sqrt{2}G_F n_{\nu_e}$ as a metric of the self-interaction potential. There are two branches, one of which has a larger growth rate than the other at all wavenumbers. The faster growing mode (blue) has a peak growth rate of $\mathrm{Im}(\omega)=6.50\times10^{10}\,\mathrm{s}^{-1}$ and a wavelength of $2.20\,\mathrm{cm}$. This is still somewhat slower and longer wavelength than the fastest growing mode in a two-beam model with equivalent densities (Appendix~\ref{sec:two_beam_ffi}). The slower growing mode has a maximum growth rate of $\mathrm{Im}(\omega)=1.06\times10^{10}\,\mathrm{s}^{-1}$ at a wavelength of $4.45\,\mathrm{cm}$. We will later show that the blue curve represents the axial symmetry preserving mode and the red curve represents the axial symmetry breaking mode.

Note that some of the unstable modes have wavelengths longer than can fit on our domain (shaded gray region), but the growth rate of these modes is significantly smaller than the peak growth rate. In addition, the finite number of grid points limits the shortest possible wavelength to $0.125\,\mathrm{cm}$, or a maximum wavenumber $k/\sqrt{2}G_F n_{\nu_e}\approx 16$, so that all modes shown in Figure~\ref{fig:dispersion_relation} outside the gray shaded region are within the grid's Nyquist limit. More importantly, our finite resolution limits the number of grid cells supporting each wavelength, and we find that to resolve the instability we need many grid cells per wavelength of the fastest growing mode. We explore the resolution requirements to obtain a converged instability growth rate in Appendix~\ref{sec:convergence} and verify that the quoted results of our simulation are robust against changes to the domain size and grid resolution.

\begin{figure}
    \centering
    \includegraphics[width=\linewidth]{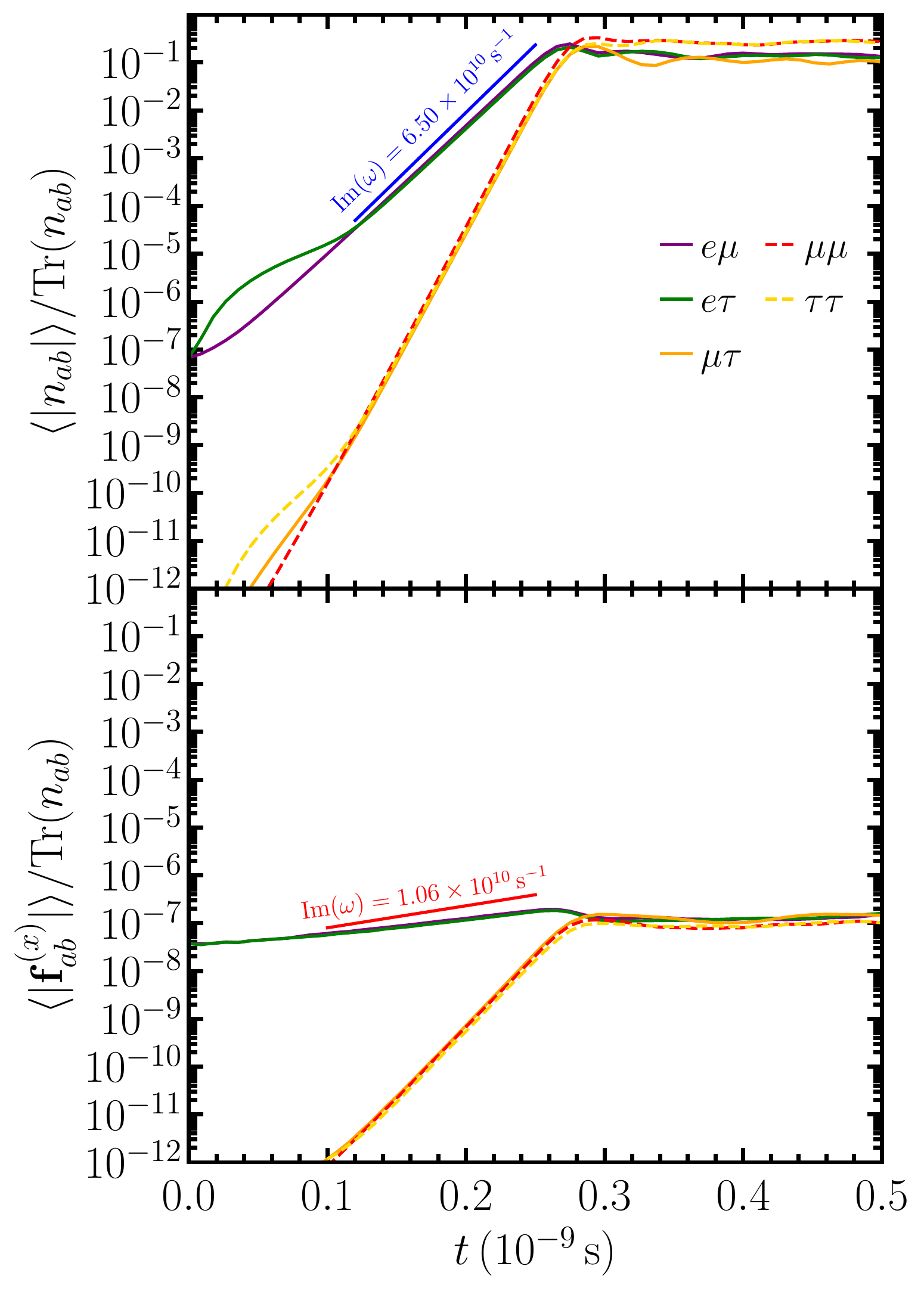}
    \caption{Time evolution of domain-averaged absolute-value of the neutrino number density (top panel) and number flux in the $\hat{x}$ direction (bottom panel). The labeled line segments show the slopes corresponding to the growth rates of the fastest growing axially symmetric (blue line segment) and axially asymmetric modes (red line segment) shown in Figure~\ref{fig:dispersion_relation}. The $e\mu$ (purple) and $e\tau$ (green) components of the number density grow at approximately the axially symmetric maximum growth rate, and the number fluxes grow at approximately the axially asymmetric growth rate. The $\mu\tau$ (orange) components grow parasitically at the same rate as the $\mu\mu$ (red dashed) and $\tau\tau$ (gold dashed) components.}
    \label{fig:1d_linear}
\end{figure}
The simulation immediately exhibits rapid growth of the flavor off-diagonal components of the neutrino density matrix. As a metric of the net growth of all modes, we calculate the magnitude of the off-diagonal components of the neutrino density and the $x$ component of the flux averaged over the entire domain as
\begin{equation}
\begin{aligned}
    \langle |n_{ab}| \rangle &= \frac{1}{N_\mathrm{zones}} \sum_\mathrm{zones}|n_{ab}|\,\,,\\
    \langle |\mathbf{f}^{(x)}_{ab}| \rangle &= \frac{1}{N_\mathrm{zones}} \sum_\mathrm{zones}\mathbf|{\mathbf{f}}^{(x)}_{ab}|\,.
    \end{aligned}
\end{equation}
$\langle |n_{ab}| \rangle$ probes the presence of modes that preserve axial symmetry, while 
$\langle |\mathbf{f}^{(x)}_{ab}| \rangle$ probes modes that break axial symmetry because 
$\langle |\mathbf{f}^{(x)}_{ab}| \rangle = 0$ would result from axially symmetric modes alone. The purple and green solid curves in Figure~\ref{fig:1d_linear} show growth of the $e\mu$ and $e\tau$ components of the neutrino density. For the first $\sim 0.1\,\mathrm{ns}$ they both exhibit rapid, but not precisely exponential growth. In this phase, there are several competing modes growing at different rates, all of which are initially large enough to contribute to $\langle |n_{e\mu}|\rangle$ and $\langle |n_{e\tau}|\rangle$. After $0.1\,\mathrm{ns}$ the average amplitudes grow exponentially until they saturate at $\sim 0.27\,\mathrm{ns}$. After saturation, the average flavor off-diagonal densities hover around the saturation level of a few tenths of the total neutrino density.

The top panel of the center column of Figure~\ref{fig:1d_diag} shows the flavor diagonal components of the neutrino density at the end of the linear phase. Electron neutrinos are still dominant, though there is now a significant fraction of $\nu_\mu$ and $\nu_\tau$, showing that the instability indeed leads to significant flavor transformation. The amplitude of the $x$-component of the flux has grown as well, though it remains at only $\sim10^{-7}$ of the total neutrino density. Although during the linear growth phase the anti-neutrino curves (dashed) match the neutrino curves (solid), as the simulation approaches saturation the nonlinear interaction between modes drives chaotic behavior and the neutrino and antineutrino distributions begin to show slight differences. The center column of Figure~\ref{fig:1d_offdiag_mag} correspondingly shows the magnitude of the flavor off-diagonal components of the neutrino density and flux. By this point in time, the off-diagonal elements have become comparable in magnitude to the diagonal components, again indicating that saturation is approaching.

The center column of Figure~\ref{fig:1d_offdiag_phase} most clearly shows that the wavelength of the growing modes matches those predicted in the dispersion relation. The top panel shows the complex phase $\phi$ of the flavor off-diagonal components of the neutrino density. The complex phase of $n_{e\mu}$ and $n_{e\tau}$ increase with $z$ and have a wavelength (distance between points of equal phase) consistent with the $2.20\,\mathrm{cm}$ wavelength of the fastest growing mode on the blue branch in Figure~\ref{fig:dispersion_relation}. The antineutrinos (dashed) are similar, except that they have a negative wavenumber, resulting in phase decreasing with increasing $z$. The bottom panel shows the complex phase of the flavor off-diagonal $x$ components of the neutrino flux. Unlike the number density, this quantity is sensitive to axial symmetry breaking and is not sensitive to modes that preserve axial symmetry. The wavelength is consistent with the $4.45\,\mathrm{cm}$ prediction from the peak of the red curve in Figure~\ref{fig:dispersion_relation}. Thus, we expect that the blue curve in Figure~\ref{fig:dispersion_relation} represents the symmetry-preserving mode and the red curve represents the symmetry-breaking mode.

The mode identification is further corroborated by the growth rates of both modes. The growth rate of the flavor off-diagonal number density (solid purple and green curves) and $x$-flux (dashed purple and green curves) in Figure~\ref{fig:1d_linear} are consistent with the growth rates predicted by the peaks of the blue and red curves, respectively, shown in Figure~\ref{fig:dispersion_relation}. The simulations do, however, show a slightly slower growth rate than predicted by the peak of the dispersion relation. This may be partially due to the fact that the domain-averaged quantities reflect many modes, all but the fastest of which grow slower than the fastest growing mode. However, the simulated growth rate also does depend somewhat on numerical considerations. We perform a convergence study in Appendix~\ref{sec:convergence} to show this dependence.

The $\mu-\tau$ mixing is more difficult to interpret. The $n_{\mu\tau}$ component (solid yellow curve in Figure~\ref{fig:1d_linear}) grows with twice the growth rate of $n_{e\mu}$ and $n_{e\tau}$. We expect that this simply follows the growth of the $n_{\mu\mu}$ (brown) and $n_{\tau\tau}$ (salmon) elements, each of which also rise with this doubled growth rate. This can be understood intuitively in terms of a rotating flavor vector $\vec{\rho}$ in SU(3) flavor space. The eight components of this vector are
\begin{equation}
\label{eq:gell-mann}
    \rho_i = \frac{1}{2} \lambda_{ab}^i \rho_{ab}\,\,,
\end{equation}
where $\lambda_{ab}^i$ are Gell-Mann matrices and $i$ is the vector index. Flavor transformation rotates this vector, preserving $\mathrm{Tr}(\rho_{ab})=1$ with constant $|\vec{\rho}|$. In this case, $|\vec{\rho}|=1/\sqrt{3}$ because we start in a nearly pure electron flavor state in Equation~\ref{eq:initial_flavor_state}. Combining these two conditions yields a relationship
\begin{equation}
    |\rho_{e\mu}|^2+|\rho_{e\tau}|^2+|\rho_{\mu\tau}|^2 = (\rho_{\tau\tau}+\rho_{\mu\mu}) - (\rho_{\mu\mu}^2 + \rho_{\tau\tau}^2 + \rho_{\mu\mu}\rho_{\tau\tau}) \,\,.
\end{equation}
If, as in the linear growth phase of the simulation, the neutrinos are only slightly perturbed from the electron flavor state, the second term on the right hand side is much smaller than the first. Additionally, $\rho_{\mu\tau}$ is much smaller than $\rho_{e\mu}$ and $\rho_{e\tau}$ throughout the linear growth phase. Combining these approximations,
\begin{equation}
    \eta \equiv \rho_{xx} - \rho_{ex}^2  \approx 0\,\,,
\end{equation}
where $\rho_{ex}^2=\rho_{e\mu}^2+\rho_{e\tau}^2$ and $\rho_{xx}=\rho_{\mu\mu}+\rho_{\tau\tau}$. Thus, if $\rho_{ex}\propto \mathrm{exp}(\mathrm{Im}(\omega)t)$ (where $x$ refers to either $\mu$ or $\tau$) then $\rho_{xx}\propto\mathrm{exp}(2\cdot \mathrm{Im}(\omega)t)$. Since the growth of $\rho_{\mu\tau}$ requires non-zero difference between $\rho_{\mu\mu}$ and $\rho_{\tau\tau}$, the growth of the initial random differences between $\rho_{\mu\mu}$ and $\rho_{\tau\tau}$ give $\rho_{\mu\tau}$ the room to grow at the same rate as $\rho_{\mu\mu}$ and $\rho_{\tau\tau}$. Note that this shows that the growth of $n_{\mu\tau}$ can be up to twice that of $n_{e\mu}$ and $n_{e\tau}$ but does not show that it must be so. It also does not explain the rapid growth of all fluxes except $\mathbf{f}_{e\mu}$ and $\mathbf{f}_{e\tau}$. More work is required for a complete description of the growth of all modes and flavors (though see \cite{chakraborty_three_2020} for work in this direction).

\begin{figure}
    \centering
    \includegraphics[width=.99\linewidth]{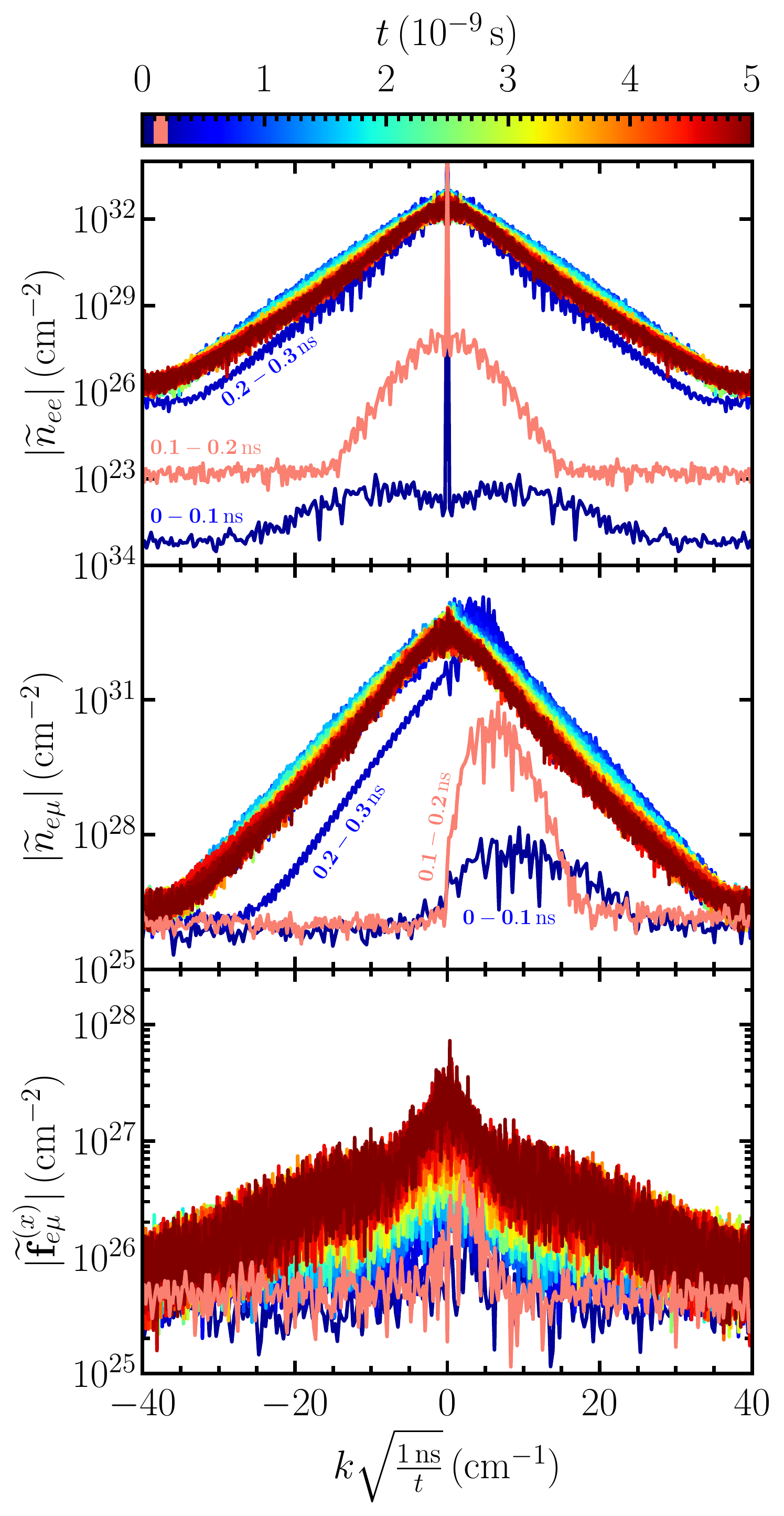}
    \caption{Fourier transforms of $n_{ee}$ (top panel), $n_{e\mu}$ (middle panel), and $\mathbf{f}_{e\mu}^{(x)}$ (bottom panel), each time-averaged over $0.1\,\mathrm{ns}$ intervals and colored by the final time of the corresponding interval. The second curve in each plot is salmon colored to highlight the time interval squarely within the linear growth phase ($0.1-0.2\,\mathrm{ns}$). The $\tilde{n}_{ee}$, $\tilde{n}_{e\mu}$, and $\tilde{\mathbf{f}}_{e\mu}^{(x)}$ pre-saturation peaks align well with the wavenumber/wavelength locations predicted in Figure~\ref{fig:dispersion_relation}, and spread out at late times to form a chaotic mixture of many modes.}
    \label{fig:fft}
\end{figure}
The differing growth rates of modes of different wavelengths can be seen in the evolution of the Fourier transforms of neutrino distribution shown in Figure~\ref{fig:fft}. Each curve in the plot is the Fourier transform of a grid quantity time-averaged over a $0.1\,\mathrm{ns}$ interval to reduce noise in the plot. The color indicates the end of the time interval for which the transform was evaluated. Beginning with the middle panel, the bottom-most curve shows the Fourier transform of $n_{e\mu}$ from $0-0.1\,\mathrm{ns}$ and the next curve (salmon) shows the same for the interval $0.1-0.2\,\mathrm{ns}$. We use salmon for the second curve to highlight the time interval squarely within the linear growth phase. The wavenumber of the peak, the minimum unstable wavenumber, and the maximum unstable wavenumber are at the locations predicted by the blue curve in Figure~\ref{fig:dispersion_relation}, as both of these intervals lie within the linear growth phase of the instability. The third curve shows the time interval $0.2-0.3\,\mathrm{ns}$, during which the instability saturates. The peak location is consistent with the wavenumber of the fastest growing modes, but additional features develop away from the peak that we will discuss in Section~\ref{sec:saturation}. The top panel of the figure shows the Fourier transform of $n_{ee}$ on the same time intervals. Since this component is purely real, the Fourier transform is symmetric in $k$, but still reflects the characteristic wavelengths in the unstable modes. The bottom panel shows the Fourier transform of $\mathbf{f}_{e\mu}^{(x)}$. Once again, the location and width of the peak in the salmon curve matches the expectations from the symmetry-breaking (red) branch of Figure~\ref{fig:dispersion_relation}.

\begin{figure}
    \centering
    \includegraphics[width=\linewidth]{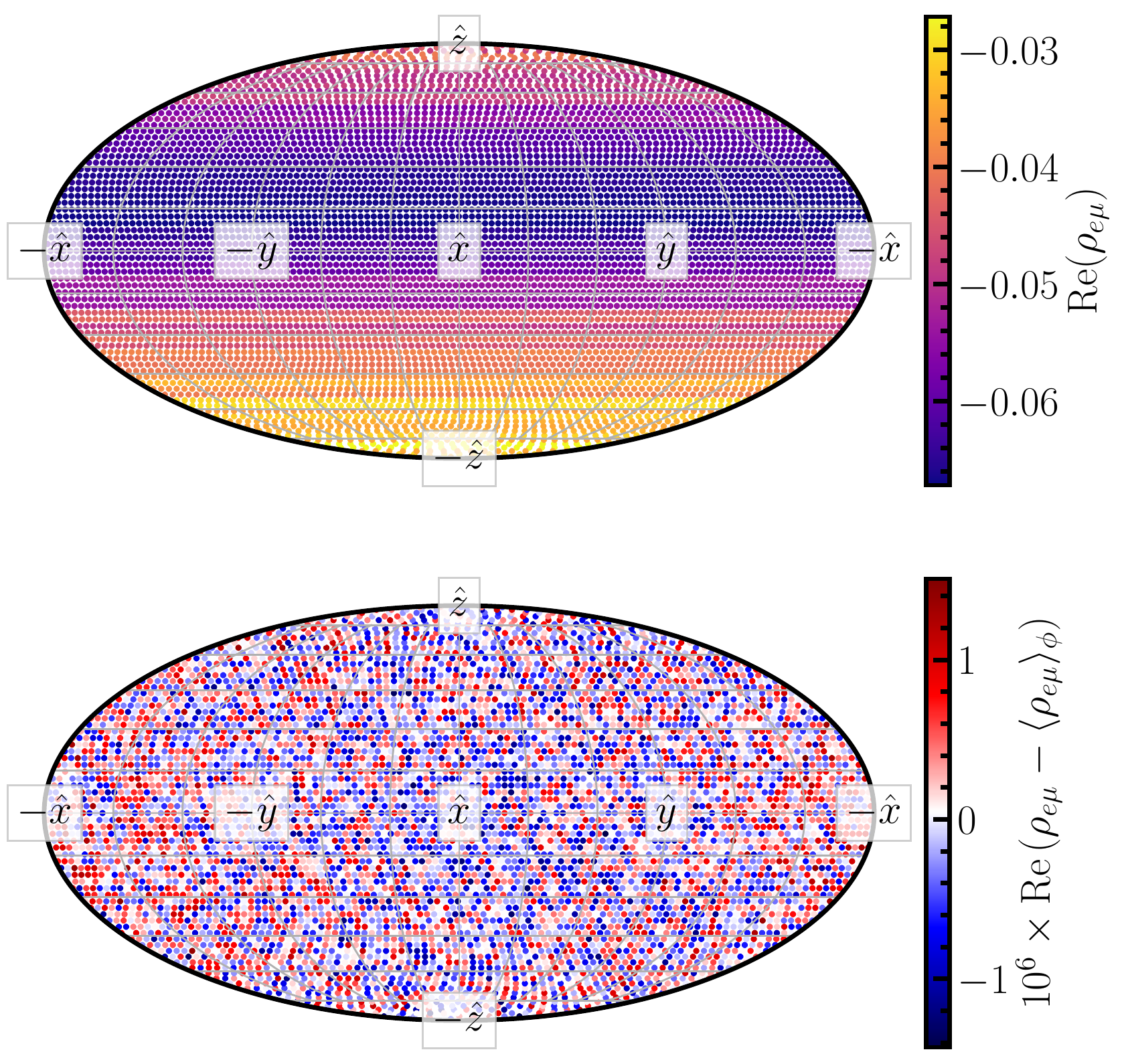}
    \caption{Mollweide projection of the neutrino particle directions in a single grid cell near the end of the linear growth phase at $t=0.27\,\mathrm{ns}$ in the ``128 Equatorial Direction'' simulation in Appendix~\ref{sec:convergence}. \textit{Top panel:} the color of each point encodes the value of $\mathrm{Re}(\rho_{e\mu})$ for each particle and shows the large amplitude of the fastest growing axially symmetric mode. \textit{Bottom panel:} the color encodes the deviation of $\mathrm{Re}(\rho_{e\mu})$ from the axial average to emphasize the axial symmetry breaking mode. The axially asymmetric mode has an amplitude comparable to the random perturbations of the initial conditions, seen as an overabundance of blue in the $\hat{x}$ direction and an overabundance of red in the $-\hat{x}$ direction.}
    \label{fig:sphere_growth}
\end{figure}
The symmetry breaking and preserving modes are also apparent in the angular structure of the neutrino flavor. The top panel of Figure~\ref{fig:sphere_growth} shows a Mollweide projection of the directions of all neutrinos located within a single grid cell for a simulation with high angular resolution. Each point on the plot represents the direction of an individual computational particle. In this high-resolution simulation we request 128 directions in the $\hat{x}-\hat{y}$ plane, visible as 128 points along the equator in the plot, for a total of 6022 particles per cell. Our fiducial simulation uses only 16 points along the equator, but we use the high angular resolution data for better visualization. We show in Appendix~\ref{sec:convergence} that the growth rates and saturation states are the same in both simulations. The colors of the points show the real part of $\rho_{e\mu}$ for each packet at $t=0.27\,\mathrm{ns}$. Although the FFI has greatly amplified these numbers from their initially random values of $\mathcal{O}(10^{-6})$, the flavor structure is very axially symmetric. If we subtract the axial average (bottom panel), more structure becomes apparent. What remains at this particular location is a slight excess of positive values in the $-\hat{x}$ direction and a slight excess of negative values in the $\hat{x}$ direction, corresponding to a small but significant flux of off-diagonal flavor $\mathbf{f}^{(x)}_{e\mu}=-1.62\times10^{24}\hat{x}\,\mathrm{cm}^{-3}$. The axial asymmetry is very weak because the growth rate of the axial symmetry breaking mode is so much slower. By the time the fastest growing axially symmetric mode saturates, the axially asymmetric mode has only grown by a factor of $\mathrm{exp}(\mathrm{Im(\omega_\mathrm{max})}\Delta t)\approx20$.

Overall, the simulations confirm the predictions of the linear growth phase from linear stability analysis. Modes of the expected wavenumbers all grow at the expected rates. The modes with the fastest growth rates appear to be the most important, as they saturate and lead to nonlinear evolution long before the slower modes can build significant amplitude.

\subsection{Saturated Phase}
\label{sec:saturation}
At $t\approx 0.27\,\mathrm{ns}$ the exponential growth of both the axially symmetric and asymmetric modes shown in Figure~\ref{fig:1d_linear} cease as the instability saturates. After saturation, the number density of all three neutrino flavors tends toward equipartition at late times. The right column of Figure~\ref{fig:1d_diag} shows a snapshot of the flavor diagonal elements of $n_{ab}$ (top) and $\mathbf{f}^{(x)}_{ab}$ (bottom) at $t=5\,\mathrm{ns}$. In both cases, the flavor content is widely variable in $z$, but when averaged over the domain, the neutrino density and flux are very close to flavor equipartition. The right column of Figure~\ref{fig:1d_offdiag_mag} shows that the flavor off-diagonal components of the density and flux remain large compared to the flavor diagonal components. Finally, the right column of Figure~\ref{fig:1d_offdiag_phase} shows that the phase is effectively randomized in the saturation phase, indicating that the coherent modes that grew in the linear phase are all but disrupted.

Figures~\ref{fig:1d_diag}-\ref{fig:1d_offdiag_phase} also show variation on much shorter wavelengths (higher wave numbers) than the fastest growing modes that dominated the linear growth phase. This evolution to short wavelengths can be seen in Figure~\ref{fig:fft}, which shows Fourier transforms of $n_{ee}$, $n_{e\mu}$ and $\mathbf{f}_{e\mu}^{(x)}$ from the $N_z=2048$ simulation in Appendix~\ref{sec:convergence}. We display this higher-resolution calculation because it allows us to track the evolution out to larger wavenumbers than our fiducial simulation, while yielding the same instability growth rates and final averaged neutrino flavor abundances as in the fiducial simulation. We described the amplification of the unstable modes shown in the earliest two curves (the lowest purple curve and the salmon curve) in Section~\ref{sec:linearGrowth}. By $0.3\,\mathrm{ns}$ (third curve from the bottom of each panel, though it is not clearly visible in the lowest panel) the amplitudes of the fastest growing modes have reached their maximal values and power in all three quantities begins to spread to larger and smaller wavenumbers. Though the noise in the flux spectra (bottom panel) makes them difficult to interpret, each curve of $|\widetilde{n}_{ee}|$ and $|\widetilde{n}_{e\mu}|$ varies linearly with 
$|k| \sqrt{1\, \mathrm{ns}/t}$ in the logarithmic plot, indicating that 
\begin{equation}
\label{eq:fft_diffusion}
    |\widetilde{n}_{ee}|\sim |\widetilde{n}_{e\mu}|\sim\mathrm{exp}\left[-|k|\left(\frac{t}{10^{-9}\,\mathrm{s}}\right)^{-1/2} \right]\,\,.
\end{equation}

Some understanding of the scaling with time and wavenumber of the post-saturation spectrum can be obtained by Fourier transforming the QKEs in Equation~\ref{eq:qke}. Again ignoring the collision term and only considering the neutrino potential, this becomes
\begin{equation}
\label{eq:qke_fft}
\begin{aligned}
    \frac{\partial \widetilde{f}_k}{\partial t} + i c \widetilde{f}_k\mathbf{\Omega}\cdot \mathbf{k} &= \frac{-i\sqrt{2}G_F}{\hbar c^3 \sqrt{2\pi}}\int d\mathbf{\Omega}' \int d\left(\frac{\nu'^3}{3}\right)(1-\mathbf{\Omega}\cdot\mathbf{\Omega}')\\
    &\int dx\, e^{-i k x}\left[f'-\bar{f}', f\right]
    \end{aligned}\,\,,
\end{equation}
where we use the convention that $\widetilde{f}_k = \int_{-\infty}^\infty dx\,e^{-i k x} f(x)/\sqrt{2\pi}$ and denote $f'=f(\mathbf{x},t,\mathbf{\Omega}',\nu')$. The primed quantities are integrated over; a neutrino with direction $\mathbf{\Omega}$ and frequency $\nu$ is moving through a distribution of neutrinos, each of which has a particular value of $\mathbf{\Omega}'$ and $\nu'$. The second term on the left is the advection term that only changes the phase of each Fourier component as waves propagate, and hence it does not affect the magnitude of the component. Representing $f'$ in terms of its Fourier components $f'=\int_{-\infty}^\infty dk'\,e^{ik' x} \widetilde{f}'_{k'}/\sqrt{2\pi}$, writing for simplicity $\widetilde{A}'_{k'}=\widetilde{f}'_{k'}-\widetilde{\bar{f}'}_{k'}$ and rearranging the integrals, the term on the bottom line of Equation~\ref{eq:qke_fft} becomes
\begin{equation}
\int_0^\infty dk'\,\left(\left[\widetilde{A}'_{k'}, \widetilde{f}_{k-k'}\right]+\left[\widetilde{A}'_{-k'}, \widetilde{f}_{k+k'}\right]\right)\,\,.
\end{equation}
Next we note that after saturation the distribution is approximately symmetric in $k$, making $\widetilde{A}'_{k'}\approx\widetilde{A}'_{-k'}$. In addition, $\widetilde{f}_k$ is strongly peaked near $k=0$. Taking inspiration from \cite{bhattacharyya_fast_2020}, this allows us to approximate $\widetilde{f}_{k-k'}+\widetilde{f}_{k+k'}\approx2\widetilde{f}_{k}+k'^2\partial_k^2 \widetilde{f}_k$. The integral can now be expressed as
\begin{equation}
\label{eq:diffusion_rhs}
    \int_0^\infty dk'\left[\widetilde{A}'_{k'},\left(2\widetilde{f}_{k}+k'^2\partial_k^2 \widetilde{f}_k\right)\right]\,\,.
\end{equation}
Similar to the behavior of the angular power spectrum in \cite{bhattacharyya_fast_2020}, the behavior of the second term is invariant to transformations where $t\rightarrow \alpha^2 t$ and $k \rightarrow \alpha k$, matching what we see at late times. Note that we do not explain why the distribution migrates from the fastest growing wavenumber to be centered at $k=0$, and we use the fact that the distribution is so sharply peaked in deriving Expression~\ref{eq:diffusion_rhs}. We leave a more detailed investigation of the saturation and decoherence of spectral power to future work.

By $t=5\,\mathrm{ns}$ the base of the wings in Figure~\ref{fig:fft} have reached a wavenumber of $|k|\approx100\,\mathrm{cm}^{-1}$, which is the maximum representable wavenumber on our grid of 2048 spatial points. Power begins to artificially build up at these higher wavenumbers, resulting in the slight upswing in the final (red) curves at $k\sqrt{1\,\mathrm{ns}/t}\approx \pm 40$. In the lower resolution fiducial simulation this upswing occurs after $t\approx2\,\mathrm{ns}$ when the wings reach that grid's maximum $|k|$ value of $50\,\mathrm{cm}^{-1}$. This suggests that in nature a true steady state solution will only be achieved by waiting long enough for the power to diffuse from the unstable wavelength to the inter-particle distance (representing the maximum physical wavenumber) multiple times, such that the distribution in wavenumber can equilibrate. However, the process of kinematic decoherence likely behaves differently in nature's three dimensions than in our simulation's single spatial dimension, so the decoherence rates demonstrated in this simulation are not necessarily realistic.

\begin{figure}
    \centering
    \includegraphics[width=\linewidth]{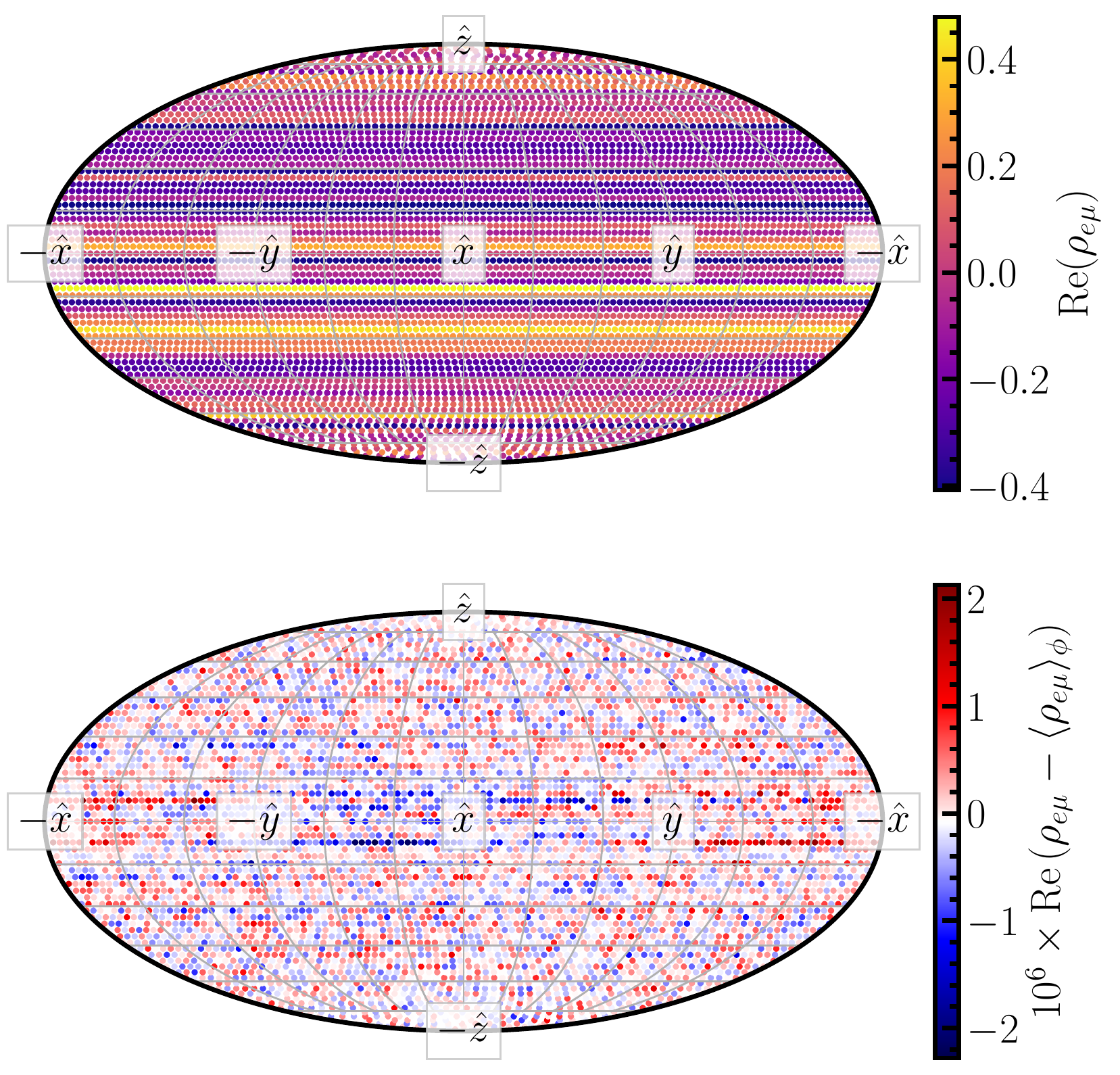}
    \caption{Mollweide projection of the neutrino particle directions in a single grid cell during the saturated phase at $t=4.95\,\mathrm{ns}$ in the ``128 Equatorial Direction'' simulation in Appendix~\ref{sec:convergence}. \textit{Top panel:} the color of each point encodes the value of $\mathrm{Re}(\rho_{e\mu})$ for that particle and shows that even in the saturated state the distribution remains largely axially symmetric. \textit{Bottom panel:} the color encodes the deviation of $\mathrm{Re}(\rho_{e\mu})$ from the axial average to emphasize that the amplitudes of axial symmetry breaking modes remain small.}
    \label{fig:sphere_saturated}
\end{figure}
Finally, we show the angular structure of the post-saturation distribution in Figure~\ref{fig:sphere_saturated} corresponding to $t=4.95\,\mathrm{ns}$. Similarly to Figure~\ref{fig:sphere_growth}, the top panel of Figure~\ref{fig:sphere_saturated} shows $\mathrm{Re}(\rho_{e\mu})$ for each particle in a single grid cell, where particle directions are plotted using a Mollweide projection. Amazingly, the saturated distribution remains highly axially symmetric, with deviations from the axial average (shown in the bottom panel) only slightly larger than they were at the end of the linear growth phase. This can be understood from the dashed curves in Figure~\ref{fig:1d_linear} that show the growth of axially asymmetric modes. Although these modes are initially unstable, the axially symmetric modes are much faster and saturate while the asymmetric modes still have small amplitude. This saturation seems not only to stop the growth of the symmetric modes, but also the asymmetric modes. However, this picture also likely changes in three spatial dimensions, so three dimensional simulations will be required for a more complete understanding of axial symmetry breaking. In addition, because the initial conditions in this simulation are themselves axially symmetric, the presence of asymmetries in the initial conditions would likely make this process rich in phenomena not expressed in our fiducial simulation.

In addition to the similarity to the angular power spectrum in Ref.~\cite{bhattacharyya_fast_2020}, these results also bear a striking resemblance to the evolution of the power spectrum in Ref.~\cite{martin_nonlinear_2019}. In the latter case, the simulations were performed in the regime of slow mode flavor transformations (i.e., the matter potential and the neutrino self-interaction potential are both relevant), but the nonlinear interaction between modes still causes power to cascade to higher modes. This was not seen in Ref.~\cite{martin_dynamic_2020}, which shows the power remaining at low wavenumbers for the duration of the simulation. We suspect this to be an effect of the choice of neutrino distribution, the structure of the initial perturbation, and the duration of the simulation. For example, \citealt{martin_dynamic_2020} use a well-controlled localized initial perturbation and allow evolution only as long as the perturbation does not travel more than the size of their periodic domain, while we use random perturbations throughout the domain and simply verify invariance of the results for different choices of the periodic domain size (Appendix~\ref{sec:convergence}). While we expect simulations using these boundary conditions to provide valuable insight into statistical properties of the instability growth and saturation, a more complete understanding of nature's realization of neutrino quantum kinetics will require a globally resolved quantum kinetics simulation of the full supernova or neutron star merger. Given our analysis above, we expect that in any situation where the initial perturbation grows to become nonlinear, power will cascade to all angular and spatial modes, though the rate at which this happens is likely to depend on the neutrino distribution itself. It is possible that the simulations of Ref.~\cite{martin_dynamic_2020} would see the same behavior on longer timescales. Given the approximate azimuthal symmetry of the fiducial simulation, we suspect that all three codes would produce statistically consistent results for the same simulation parameters, though this remains to be checked. However, initial conditions and perturbations that break the azimuthal symmetry in Section~\ref{sec:pstudy} will require a simulation method that simulates neutrinos in the relevant number of dimensions.

\section{Parameter Study}
\label{sec:pstudy}
Our final goal in this work is to provide a first look into the dependence of the FFI on the initial matter and neutrino distributions. The initial conditions described in Section~\ref{sec:initialConditions} rely on nine input parameters: the number density of neutrinos $n_{\mathrm{input}}$ and antineutrinos $\bar{n}_{\mathrm{input}}$, the number flux vector of neutrinos $\mathbf{f}_{\mathrm{input}}$ and antineutrinos $\bar{\mathbf{f}}_{\mathrm{input}}$ (three components each), and the background electron density $n_e$. Although we have shown that {\tt Emu} simulates the FFI with high fidelity, it is impractical to densely sample the full parameter space, even under this limited subset of possible initial conditions. Instead, we scratch the surface here by performing a parameter sweep in a single input variable at a time, holding the other input parameters the same as in the fiducial simulation.

\begin{figure}
    \centering
    \includegraphics[width=.99\linewidth]{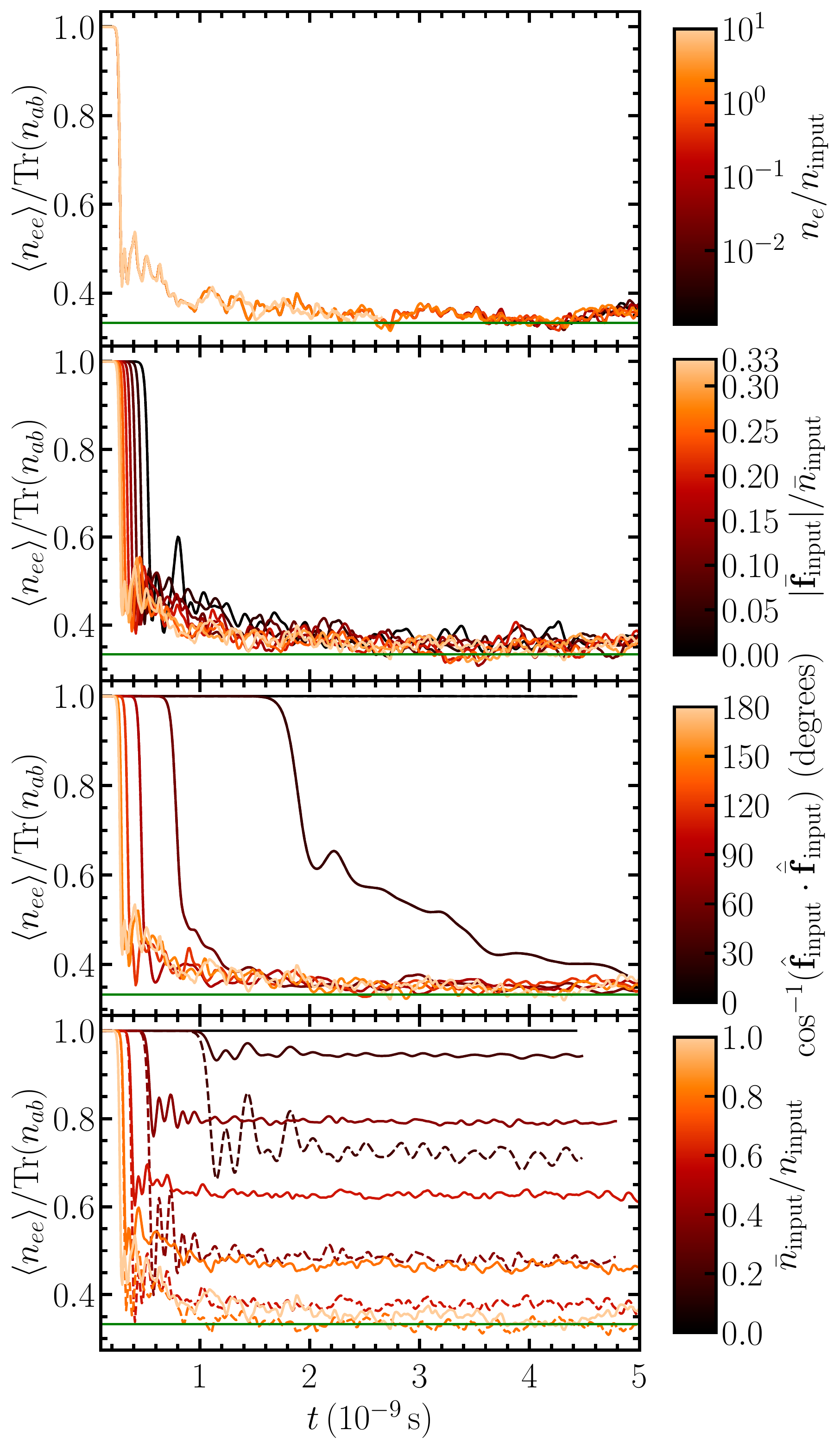}
    \caption{Time evolution of the electron diagonal component of the domain-averaged neutrino density (solid lines) and antineutrino density (dashed lines) in our parameter study. From top to bottom, the simulations shown varied the background electron density (first panel), the antineutrino flux factor (second panel), the antineutrino flux direction (third panel), and the antineutrino density (fourth panel). The color indicates the value of the input parameter that is varied from the fiducial simulation and the horizontal green line is at a value of $1/3$. The electron density affects neither the growth rate nor the saturation state of the instability. All other factors affect the growth rate. That only the antineutrino density affects the final saturation state is an artifact of the neutrino-antineutrino symmetry of the fiducial simulation.}
    \label{fig:parametersweep}
\end{figure}
The results of these simulations are summarized in Figure~\ref{fig:parametersweep}, where we plot the time evolution of the domain average abundance of electron neutrinos. Each curve is a distinct simulation with the input parameter varied from the fiducial simulation indicated by the color. In the top panel, we show the results of seven simulations identical to the fiducial simulation except that we have included a nonzero background electron density $n_e$. The fiducial simulation has $n_e=0$ and is plotted as the darkest curve, though it is occulted by the other curves. As described in Section~\ref{sec:results}, the instability saturates after $\sim0.3\,\mathrm{ns}$, resulting in a significant decrease in the number of electron neutrinos. At late times, the number density of electron neutrinos (solid curve) and antineutrinos (dashed curve) approaches one third of the number density (green line) of all neutrinos or antineutrinos, respectively, indicating flavor equilibration. As expected from the stability analysis, the electrons modify the real component of the mode frequencies but do not affect the growth rate of the unstable modes. Thus, the simulations varying $n_e$ proceed very similarly, diverging at later times due to the chaotic nature of the saturation phase.

In the second panel from the top we show a series of eight simulations where we hold the number densities and flux directions of neutrinos and antineutrinos constant, but vary the magnitude of the antineutrino flux from 0 (isotropic, black curve) to $\bar{n}_{\mathrm{input}}/3$ (fiducial simulation, light orange curve). In the third panel from the top we show results from a series of seven simulations where we hold the neutrino densities and flux magnitudes constant, but change the direction of the antineutrino flux from aligned with the neutrino flux (black curve) to opposed to the neutrino flux (fiducial simulation, light orange curve). In both cases, the parameter variation reduces the growth rate of the instability, increasing the time when the instability saturates. However, these parameters do not appear to change the final abundances of neutrino flavors. All simulations approach flavor equilibration except for the case where the antineutrino and neutrino fluxes align and no flavor transformation occurs. Also note that when we vary either the magnitude or direction of antineutrino flux, the neutrinos and antineutrinos follow the same trajectory and are plotted one on top of the other.

Finally, in the bottom panel we show the results of six simulations where we vary the number density of antineutrinos, keeping the direction and flux factor ($|\bar{\mathbf{f}}_{\mathrm{input}}|/\bar{n}_{\mathrm{input}}=1/3$) constant. The fiducial simulation has $\bar{n}_{\mathrm{input}}/n_{\mathrm{input}}=1$ and is plotted as the lightest orange curve. As we decrease the number of electron antineutrinos in the initial conditions (moving toward darker colors), the growth rate of the instability slows, indicated by a longer time before the instability saturates. In addition, the final abundances of both neutrinos and antineutrinos increase relative to the fiducial simulation, but neutrinos increase at a more rapid rate than antineutrinos. When there are no antineutrinos present (uppermost black line) the distribution shows no flavor transformation.

\begin{figure}
    \centering
    \includegraphics[width=\linewidth]{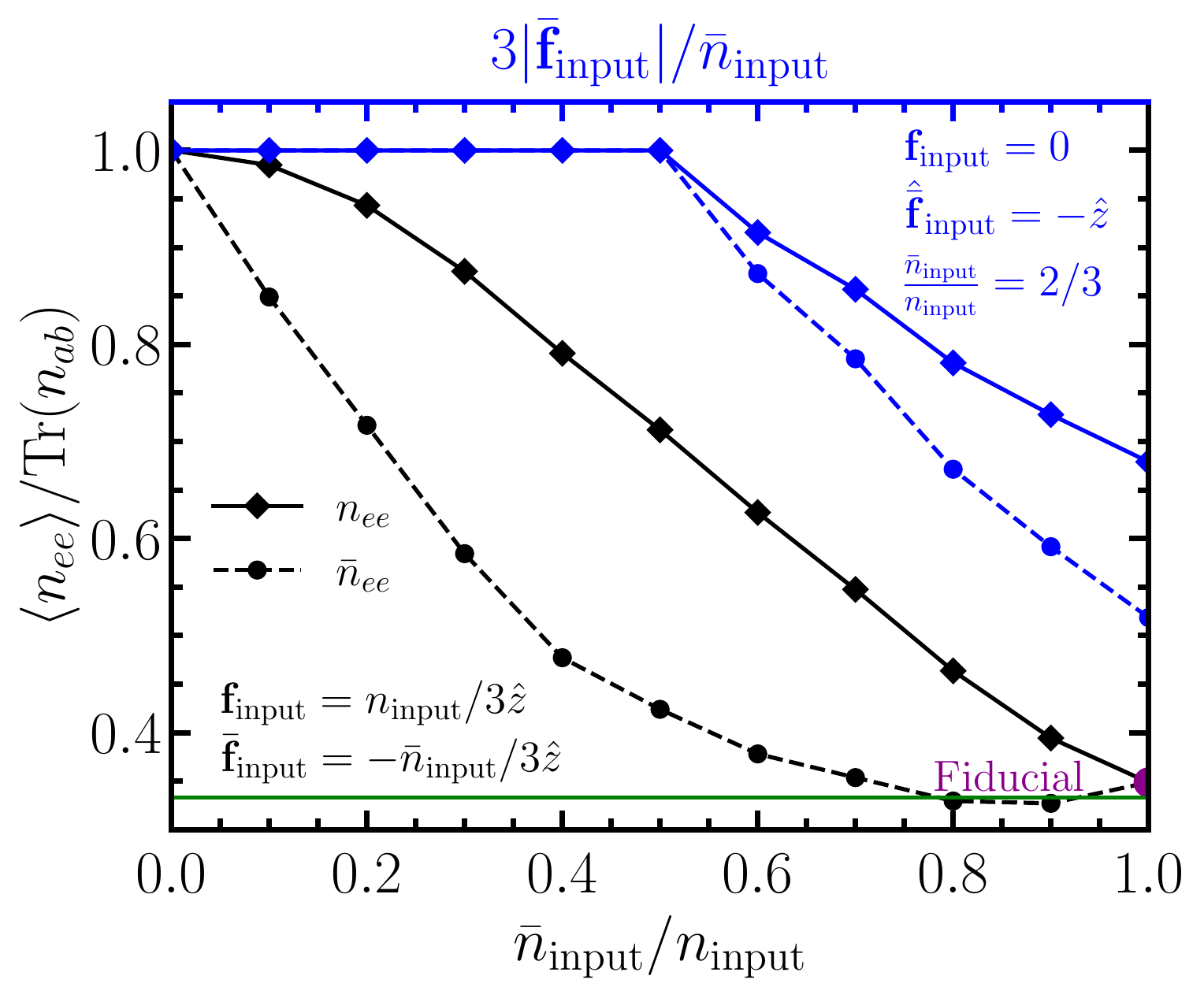}
    \caption{Time averages of the electron neutrino (solid) and antineutrino (dashed) abundances in the post-saturation phase for parameter sweeps of $|\bar{\mathbf{f}}_{\mathrm{input}}|$ and $\bar{n}_{\mathrm{input}}/n_{\mathrm{input}}$, run for $25\,\mathrm{ns}$. In blue, we show the results of a set of simulations with fixed $\bar{n}_{\mathrm{input}}/n_{\mathrm{input}}$, $\mathbf{f}_{\mathrm{input}}$, and $\hat{\bar{\mathbf{f}}}_{\mathrm{input}}$ and vary $|\bar{\mathbf{f}}_{\mathrm{input}}|$ from 0 to $\bar{n}_{\mathrm{input}}/3$. The distribution is unstable only when there is an ELN crossing ($3|\bar{\mathbf{f}}_{\mathrm{input}}|/\bar{n}_{\mathrm{input}}> 1/2$). In black, we show the results of simulations where we set $\mathbf{f}_{\mathrm{input}}$ and $\bar{\mathbf{f}}_{\mathrm{input}}$ and vary $\bar{n}_{\mathrm{input}}/n_{\mathrm{input}}$ from $0 - 1$ (the same data as in the bottom panel of Figure~\ref{fig:parametersweep}). The fiducial simulation is plotted as a purple point, and the horizontal green line marks complete flavor equilibration ($\langle n_{ee} \rangle=\mathrm{Tr}(n_{ab})/3$). In both parameter sweeps, the antineutrinos consistently experience a greater degree of relative flavor transformation.}
    \label{fig:parametersweep_averages}
\end{figure}

However, the fact that only the ratio of antineutrino to neutrino density changes the final flavor abundances is an artifact of our choice of initial conditions for our fiducial simulation. We performed another set of simulations with initial parameters $\bar{n}_{\mathrm{input}}/n_{\mathrm{input}}=2/3$, $\mathbf{f}_{\mathrm{input}}= 0$, and $\hat{\bar{\mathbf{f}}}_{\mathrm{input}}=-\hat{z}$ and varied $|\bar{\mathbf{f}}_{\mathrm{input}}|$ from 0 to $\bar{n}_{\mathrm{input}}/3$. Using these parameters, there is an ELN crossing only for $3|\bar{\mathbf{f}}_{\mathrm{input}}|/\bar{n}_{\mathrm{input}}\ge 1/2$. We run each simulation for $25\,\mathrm{ns}$ to ensure that all unstable simulations had sufficient time to reach an approximately stationary final state. In Figure~\ref{fig:parametersweep_averages} we plot time averages of the electron neutrino (solid) and antineutrino (dashed) abundances in blue for the post-saturation phase. As expected, when there is no ELN crossing ($3|\bar{\mathbf{f}}_{\mathrm{input}}|/\bar{n}_{\mathrm{input}} < 1/2$) the distribution exhibits no flavor transformation. Beyond this point, increasing the flux factor increases the size of the ELN crossing, increases the growth rate of the instability, and increases the amount of eventual flavor transformation. For none of our simulations in this parameter sweep do we see complete flavor equilibration
($\langle n_{ee} \rangle=\mathrm{Tr}(n_{ab})/3$), though this may occur for larger antineutrino flux factors. For comparison, we also plot the same quantities for the $\bar{n}_{\mathrm{input}}/n_{\mathrm{input}}$ parameter sweep (i.e., calculated from the results shown in the bottom panel of Figure~\ref{fig:parametersweep}) in black and label the fiducial simulation in purple. In both parameter sweeps the antineutrinos consistently experience a greater degree of relative flavor transformation. This is simply a reflection of the fact that in all of our simulations, the same number of neutrinos and antineutrinos oscillate. That is, we find that in our simulations
\begin{equation}
\label{eq:neutrino_antineutrino_final_state}
    n_{ee}(0)-\langle n_{ee}\rangle = \bar{n}_{ee}(0)-\langle \bar{n}_{ee}\rangle\,\,.
\end{equation}
As elsewhere in the paper, all variations of $n$ in the equation are number densities with units $\mathrm{cm}^{-3}$, and in our simulations $n_{\mathrm{input}}$ and $n_{ee}(0)$ are equivalent. Since there are fewer antineutrinos than neutrinos, a larger fraction must change flavor to match the total number of oscillating neutrinos. This is not entirely surprising. Since the neutrino self-interaction potential dominates the total potential, neutrinos and antineutrinos experience the same potential (up to a sign) and the flavor vectors of an individual particle rotate the same amount for neutrinos and antineutrinos, though in opposite directions. Another way to say this is that the total lepton number is approximately preserved when the equations of motion are approximately matter-antimatter symmetric. This is in line with previous predictions derived assuming a negligible contribution from the vacuum potential (e.g. Equation 5 in \citealt{martin_dynamic_2020}), but in situations where the vacuum potential is relevant, the matter-antimatter symmetry of the evolution equations is broken and this relationship no longer holds.

Equation~\ref{eq:neutrino_antineutrino_final_state} does not predict the amount of neutrinos that oscillate, but if this result holds in general for fast flavor transformations, it does say that the net electron flavor content (i.e. number of electron neutrinos minus the number of antineutrinos) does not change under the fast flavor instability even in saturation. In addition, it provides a bound on the amount of neutrinos that change flavor. In the limit of zero heavy lepton neutrinos, $n_{ee}(0)-\langle n_{ee}\rangle=\mathrm{min}(n_{ee}(0),\bar{n}_{ee}(0))$, since it is not possible for more electron neutrinos to change flavor than there exist electron neutrinos. We can extend this to three flavors, noting that the amplitude of possible flavor transformation is given by $\sqrt{3}l$, where $l$ is the length of the $\mathrm{SU}(3)$ flavor isospin vector (analogous to Equation~\ref{eq:gell-mann} for $n_{ab}$ instead of $\rho_{ab}$). That is,
\begin{equation}
\label{eq:flavor_bound}
    |n_{ee}(0)-\langle n_{ee}\rangle| \leq \sqrt{3} \,\mathrm{min}(l,\bar{l})
\end{equation}
and similarly for antineutrinos. Explicitly, when the flavor off-diagonal elements are zero, these can be written
\begin{equation}
    l^2 = \frac{1}{6}\left[(n_{ee}-n_{\mu\mu})^2+(n_{ee}-n_{\tau\tau})^2+(n_{\mu\mu}-n_{\tau\tau})^2\right]
\end{equation}
and similarly for antineutrinos. If the antineutrino density flavor vector length is smaller than that of the neutrinos, the flavor rotation of the antineutrinos is not bound, but the flavor rotation of neutrinos must match. Complete flavor equilibration of both neutrinos and antineutrinos is possible only if both flavor vector lengths are equal. Once again, these bounds do not actually predict the final abundances; the abundances of electron neutrinos in Figure~\ref{fig:parametersweep_averages} show an actual amount of flavor change that is significantly smaller than the bound. For instance, the bound for the fiducial simulation only says that the $\langle n_{ee}\rangle\geq 0$.

In summary, we show that the details of the neutrino distribution affect both the growth rate of the FFI and the final abundances of each neutrino flavor. Although we do not have a general way to map initial conditions to final neutrino abundances, we show that a small ELN crossing can produce minimal flavor change. Clearly, the amount of flavor transformation present in the final saturated state must be mapped out by a more comprehensive parameter sweep. We save this for future work.

\section{Conclusions}
\label{sec:conclusion}
We present the new open-source particle-in-cell (PIC) neutrino quantum kinetics code {\tt Emu}. The PIC algorithm evolves the quantum states of individual computational particles by accumulating number density and flux on a background grid and then interpolating the net number density and flux from the background grid to evaluate the potential at each individual neutrino's location. The code is based on the AMReX framework and is performance portable to CPU and GPU hardware. We perform a series of test problems and a resolution study in the appendix.

We use {\tt Emu} to perform simulations of the neutrino fast flavor instability (FFI) in one spatial dimension and two momentum (direction) dimensions. We simulate and analyze in detail a distribution of initially electron neutrinos and antineutrinos with number fluxes in opposite directions and a flux factor of 1/3 (Figures~\ref{fig:1d_diag}-\ref{fig:1d_offdiag_phase}). We demonstrate the simultaneous evolution of the axial symmetry-preserving and symmetry-breaking modes with growth rates (Figure~\ref{fig:1d_linear}) and wavelengths (Figure~\ref{fig:1d_offdiag_phase}) that match predictions from linear stability theory (Figure~\ref{fig:dispersion_relation}). The symmetry-preserving mode has the faster growth rate, resulting in a high degree of axial symmetry during instability growth (Figure~\ref{fig:sphere_growth}) that persists into the post-saturation phase (Figure~\ref{fig:sphere_saturated}). After the instability saturates, spatial fluctuations in neutrino flavor cascade from the length scale corresponding to the wavelength of the fastest growing axial symmetry mode to high wavenumbers (Figure~\ref{fig:fft}) following a simple analytic relationship (Equation~\ref{eq:fft_diffusion}). The final abundances of neutrinos and antineutrinos in this simulation approach flavor equipartion at late times.

We then vary the initial distribution of antineutrinos from the fiducial simulation in a series of simulations. We show that that the antineutrino number density, flux magnitude, and flux direction all modify the growth rate of the instability (Figure~\ref{fig:parametersweep}). In addition, we show that varying antineutrino flux factor and number density result in different final abundances of each neutrino flavor after the instability saturates (Figure~\ref{fig:parametersweep_averages}). We furthermore show that a small ELN crossing leads to slower instability and little flavor change between the initial and post-saturation states, and that the growth rate and flavor change increase with larger ELN crossings (Figure~\ref{fig:parametersweep_averages}). Although we do not have a way of mapping the initial distributions to final flavor abundances, we do demonstrate that in all of our simulations the same number of neutrinos and antineutrinos change flavor. That is, the final abundances of neutrino flavors imply the final abundances of antineutrino flavors (Equation~\ref{eq:neutrino_antineutrino_final_state}). This relationship results in a bound on the amount of neutrino and antineutrino flavor change (Equation~\ref{eq:flavor_bound}), though the simulation results never come close to this bound.

Our initial parameter study based on our highly symmetric fiducial simulation (Figure~\ref{fig:parametersweep}) demonstrated how a sparse exploration of the parameter space can lead to incomplete conclusions. Exploring more variations of the initial conditions (blue curve in Figure~\ref{fig:parametersweep_averages}) revealed that the antineutrino flux factor affects the final abundances of neutrino flavor, even though the parameter sweep in Figure~\ref{fig:parametersweep} showed all antineutrino flux factors resulting in flavor equilibration. We expect that it will be possible to build a mapping from initial distributions to final neutrino abundances, but this will require many more simulations. One major shortcoming of this work is that the simulations in this work were performed in a single spatial dimension. Although a dense covering of the parameter space using three-dimensional simulations is not currently feasible, it will be important to anchor the 1D simulations with select 3D ones. This, too, we leave to future work.

We also expect that the one-dimensional nature of the simulations in this work will significantly affect the process of kinematic decoherence and cascade of power to small scales due to the artificially reduced number of degrees of freedom. In addition, all of the neutrinos in these simulations were of a single energy. Although we include neutrino mass terms, the neutrino potential in these simulations is much larger than the vacuum potential, so all of these simulations are well outside the realm of slow collective oscillations. Many of these conclusions likely do not apply when the vacuum potential becomes significant. This is yet another avenue for future study.

With these prospects in mind, we note that the {\tt Emu} implementation of modern and established numerical methods from the plasma physics community enables efficient numerical predictions with even modest computational resources. For scale, each of the simulations in the parameter survey in Figure~\ref{fig:parametersweep} requires about 30 minutes on an NVIDIA P100 GPU. The development of this code was greatly accelerated by open access to the open-source plasma physics code {\tt Warp-X}. This goes to show the importance of open-source software beyond the code's intended purpose. We are hopeful that the development of numerical techniques and open-source software will accelerate the field's convergence to a phenomenological theory of fast flavor transformations.

\section{Acknowledgements}
We are very grateful to Gail McLaughlin, Samuel Flynn, Evan Grohs, Revathi Jambunathan, Luke Johns, James Kneller, Hiroki Nagakura, and MacKenzie Warren  for many useful discussions. This material is based upon work supported by the National Science Foundation under Award No. 2001760. SR is supported by the N3AS Fellowship under National Science Foundation grant PHY-1630782 and Heising-Simons Foundation grant 2017-228. This research was supported by the Exascale Computing Project (17-SC-20-SC), a collaborative effort of the U.S. Department of Energy Office of Science and the National Nuclear Security Administration. This work was supported in part by the U.S. Department of Energy, Office of Science, Office of Workforce Development for Teachers and Scientists (WDTS) under the Science Undergraduate Laboratory Internship (SULI) program. This work used the Bridges system, which is supported by NSF award number ACI-1445606, at the Pittsburgh Supercomputing Center (PSC). This research used the Cori supercomputer of the National Energy Research Scientific Computing Center (NERSC), a U.S. Department of Energy Office of Science User Facility located at Lawrence Berkeley National Laboratory, operated under Contract No. DE-AC02-05CH11231.

In this work we make use of {\tt SymPy} \cite{meurer_sympy_2017}, {\tt NumPy} \cite{walt_numpy_2011}, {\tt MatPlotLib} \cite{hunter_matplotlib_2007}, {\tt SciPy} \cite{virtanen_scipy_2020}, {\tt yt} \cite{turk_yt_2011}, and {\tt Mathematica} \cite{wolfram_research_inc_mathematica_2020}.

\bibliography{references}

\appendix

\section{Test Problems}

\subsection{Mikheyev-Smirnov-Wolfenstein (MSW) Oscillations}
\label{sec:msw_test}
\begin{figure}
    \centering
    \includegraphics[width=\linewidth]{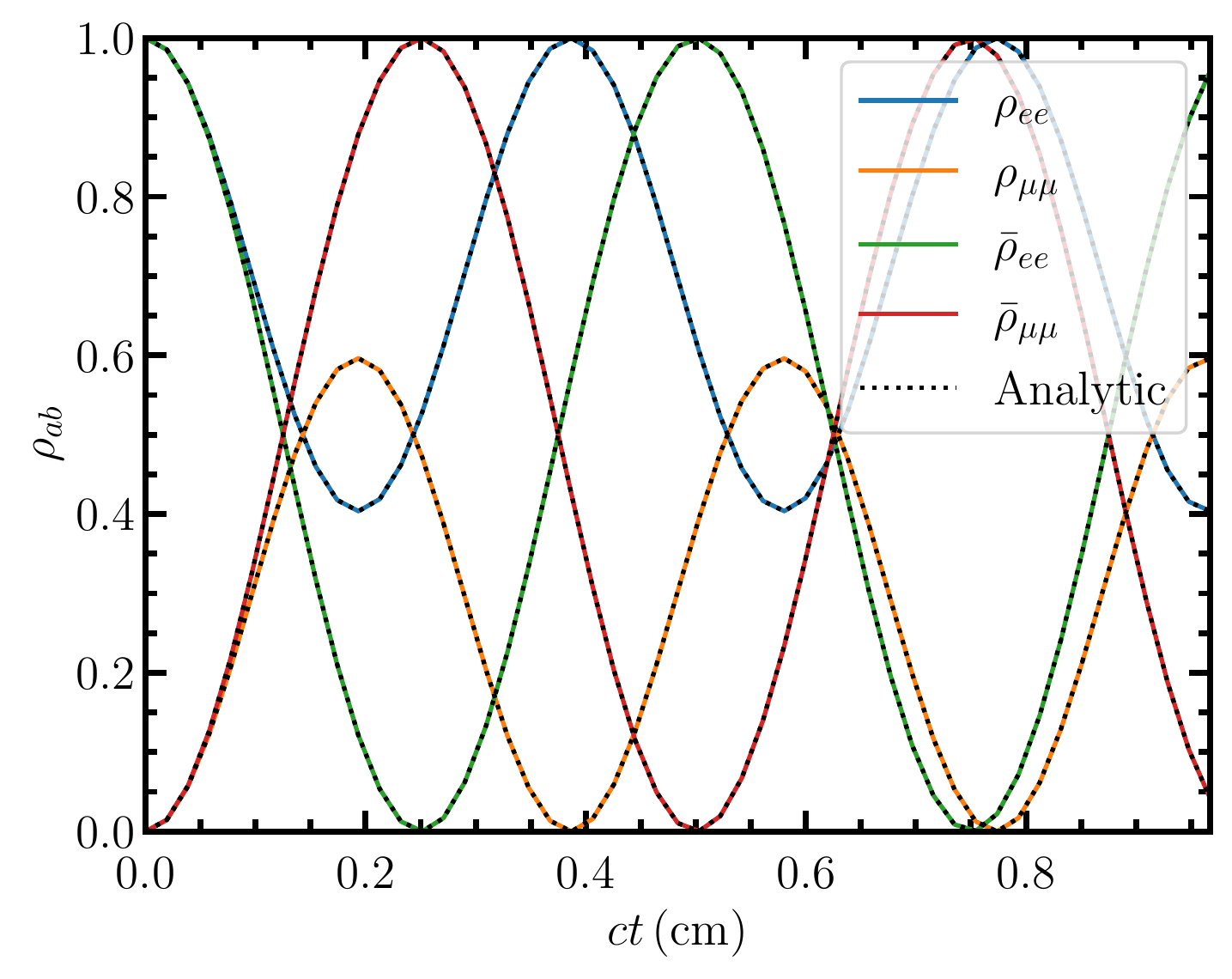}
    \caption{A two-flavor ($\nu_e$ and $\nu_\mu$) one-zone MSW test. We set the background matter, neutrino masses, and flavor mixing angles such that antineutrinos undergo complete flavor oscillations with a period of $ct=1\,\mathrm{cm}$. The density matrix of each (anti-)neutrino is shown as a function of time, with the electron flavor in blue (green) and the muon flavor in orange (red). The analytic results are overplotted as dotted black curves and show good agreement. The antineutrinos undergo a complete transformation, but since the neutrinos are not in resonance they do not completely swap flavors.}
    \label{fig:msw_test}
\end{figure}
To test our implementation of the vacuum and matter potentials we follow \cite{richers_neutrino_2019} and perform a simple two-flavor one-zone test of neutrino oscillations in a regime where the neutrino potential is negligible and the matter density is tuned to put the neutrino oscillations in resonance. Specifically, we set up a 1 cm cube in a $1\times 1\times 1$ domain filled with matter at a density of $1.35\times 10^{9}\,\mathrm{g\,cm}^{-3}$ and an artificial electron fraction of $Y_e=1$. The neutrino mass eigenstates have masses of $m_1=8.60\times10^{-3}\,\mathrm{eV}$ and $m_2=0\,\mathrm{eV}$, and the mixing angle is set to $\theta_{12}=33.82$ degrees. We only initialize two particles, each of which represents a single neutrino and a single antineutrino at an energy of $h\nu = (m_2^2-m_1^2) c^4 \sin(2\theta_{12}) / (8\pi\hbar c) = 0.138\,\mathrm{eV}$. This combination of background matter and neutrino properties make the antineutrinos undergo complete flavor oscillations with a period of $ct=1\,\mathrm{cm}$. The flavor-diagonal values of the density matrix of each neutrino and antineutrino are plotted as colored curves in Figure~\ref{fig:msw_test} and the analytic results (e.g. \cite{bellini_neutrino_2014}) are overplotted as dotted black curves. The antineutrino values (green and red) undergo complete transformation, but since the neutrinos (blue and orange) are not in resonance they do not completely swap. The maximum difference between the analytic and simulated quantities is $3.7\times10^{-4}$ when using a timestep of $6.5\times10^{-13}\,\mathrm{s}$.

\subsection{Bipolar Oscillations}
\begin{figure}
    \centering
    \includegraphics[width=\linewidth]{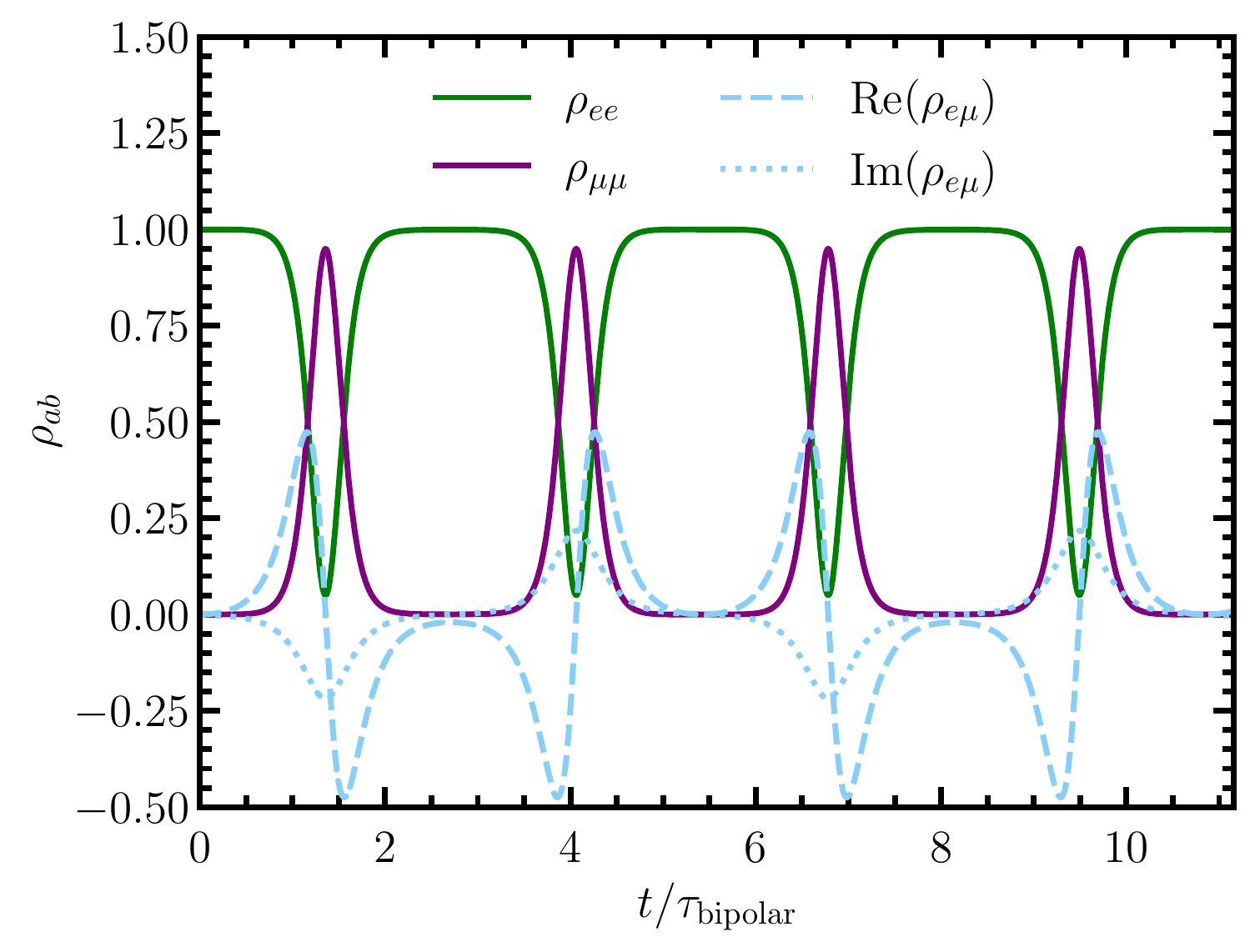}
    \caption{An isotropic one-zone bipolar oscillation test using two counter-propagating beams. We set the mixing angle $\theta_{12}$ and background matter density such that $\mathcal{H}_\mathrm{neutrino}/\mathcal{H}_\mathrm{vacuum}=10$ with characteristic oscillation time $\tau_{\mathrm{bipolar}} = 8.96\times10^{-4}\,\mathrm{s}$. The solid curves show the evolution of the diagonal components of the neutrino density matrix (electron flavor in green, muon flavor in purple). The dashed and dotted blue curves show the evolution of the real and imaginary parts of the off-diagonal element, respectively. Our results match those of \cite{hannestad_self-induced_2006} and \cite{richers_neutrino_2019}.}
    \label{fig:bipolar_test}
\end{figure}
To test our implementation of the isotropic components of the self-interaction potentials we follow \cite{richers_neutrino_2019} and perform an isotropic bipolar oscillation test. Once again, we use only one particle moving in the $+z$ direction and one moving in the $-z$ direction, since this exactly replicates the behavior of an isotropic distribution in the context of this test. We choose particle weights to ensure the neutrino and antineutrino number densities are 
$n = \bar{n} = 10 (m_2 - m_1)^2 c^4 / (2 \sqrt{2} G_F E)$ 
with energy $E=50~\mathrm{MeV}$. We use the same neutrino masses as in Section~\ref{sec:msw_test}, but we instead choose a mixing angle of $\theta_{12}=0.01=0.573\,\mathrm{degrees}$ and a background mass density of zero. This corresponds to the $\mathcal{H}_\mathrm{neutrino}/\mathcal{H}_\mathrm{vacuum}=10$ case shown in Ref.~\cite{hannestad_self-induced_2006} with a characteristic oscillation time of $\tau = 8.96\times10^{-4}\,\mathrm{s}$ (see their Equation 21). The solid curves in Figure~\ref{fig:bipolar_test} show the evolution of the diagonal components of the neutrino density matrix, while the dashed and dotted curves show the evolution of the real and imaginary parts of the off-diagonal element, respectively. We do not perform a detailed error calculation since the analytic result in Ref.~\cite{hannestad_self-induced_2006} is itself approximate, but the curves in Figure~\ref{fig:bipolar_test} match those in Ref.~\cite{hannestad_self-induced_2006} and \cite{richers_neutrino_2019}. Note that unlike \cite{richers_neutrino_2019} we use an inverted mass hierarchy, so we start in an electron flavor eigenstate instead of a muon flavor eigenstate.

\subsection{Homogeneous Two Beam Fast Flavor Instability}
\label{sec:homogeneous_two_beam_ffi}
\begin{figure}
    \centering
    \includegraphics[width=\linewidth]{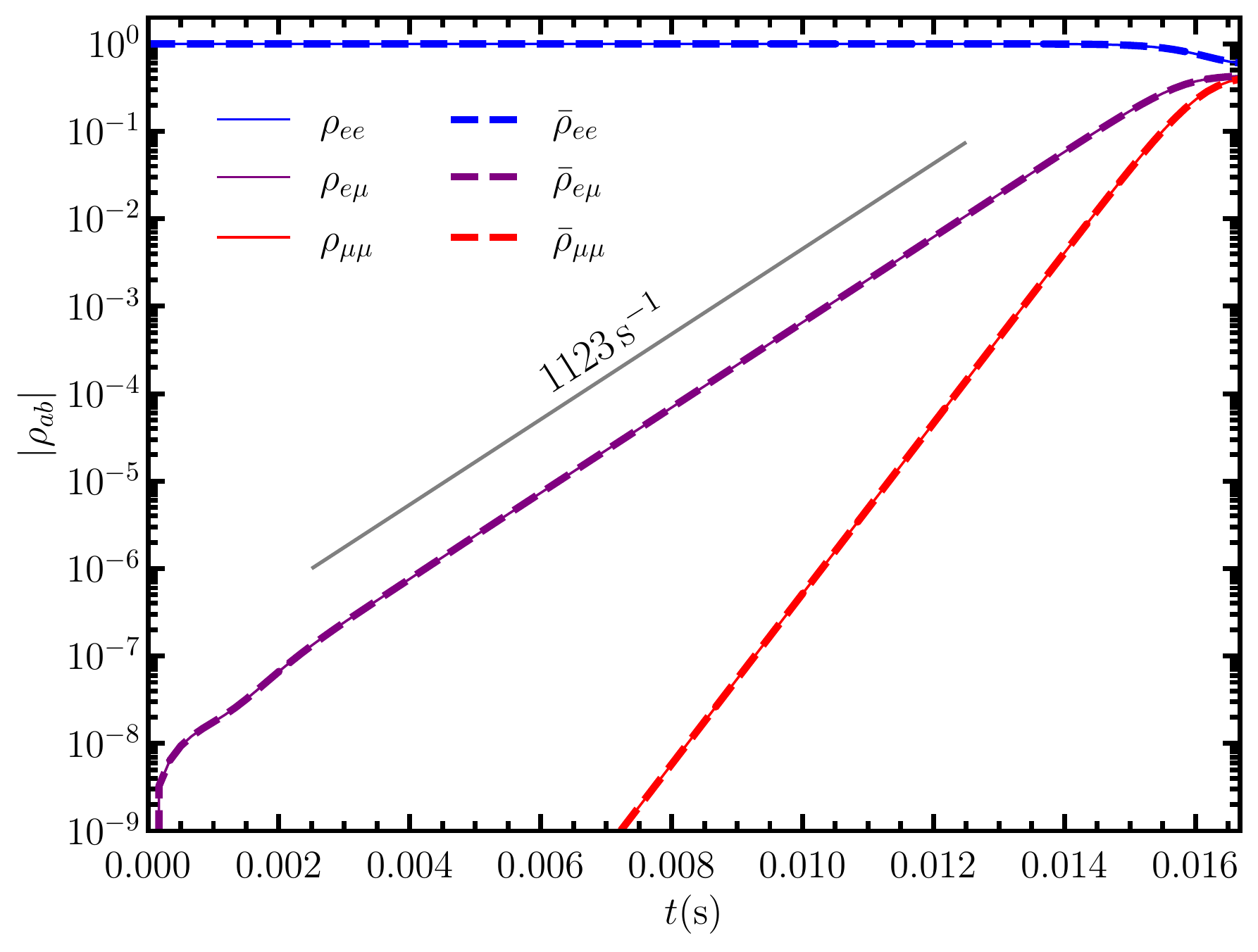}
    \caption{Homogeneous ($k=0$) two-flavor two-beam fast flavor instability prepared in an initially pure electron flavor state. The instability grows exponentially, as electron (blue) neutrinos (solid) and antineutrinos (dashed) rotate toward the pure muon state (red) and a quantum superposition of electron and muon flavor states (purple). The growth rate of the flavor off-diagonal component matches the prediction from linear stability theory well (gray line segment).}
    \label{fig:k0_fast_flavor_test}
\end{figure}
To test the anisotropic terms in the self-interaction potential, we perform a two-flavor two-beam fast flavor conversion test. The growth rate of the fast flavor instability in this simplified setup is easily calculated analytically as described by \cite{chakraborty_self-induced_2016}. We use the same neutrino masses as in Section~\ref{sec:msw_test}, but choose a mixing angle of $\theta_{12}=10^{-6}\,\mathrm{degrees}$. We use a single grid cell containing one particle moving in the $+z$ direction representing only neutrinos, one moving in the $-z$ direction representing only antineutrinos, and a vacuum matter background. The number of neutrinos each particle represents is set such that the total number density of each of neutrinos and antineutrinos is $n=(m_2^2-m_1^2) c^4/(4 \sqrt{2}G_F h\nu)=2.92\times10^{24}\,\mathrm{cm}^{-3}$ (where $G_F$ here has units of $\mathrm{erg\,cm}^3$). This yields the fastest growing instability given the neutrino parameters according to Equation 2.10 in Ref.~\cite{chakraborty_self-induced_2016}. 

The results are shown in Figure~\ref{fig:k0_fast_flavor_test}. The blue curve on top shows the evolution of the electron diagonal component of the density matrix for neutrinos (solid) and antineutrinos (dashed). As time progresses, the neutrinos mix into the the muon flavor state (red) and a quantum superposition of muon and electron flavor states (purple). Linear stability theory predicts an exponential increase in the off-diagonal component with a growth rate of $\mathrm{Im}(\omega) = (m_2^2-m_1^2)c^4/(2\hbar h\nu)=1123\,\mathrm{s}^{-1}$. The fitted growth rates between $t=8.4\,\mathrm{ms}$ and $11.7\,\mathrm{ms}$ match the theoretical rate with a relative error of at most $3.2\times10^{-5}$.

\subsection{Inhomogeneous Fast Flavor Instability}
\label{sec:two_beam_ffi}

\begin{figure}
    \centering
    \includegraphics[width=\linewidth]{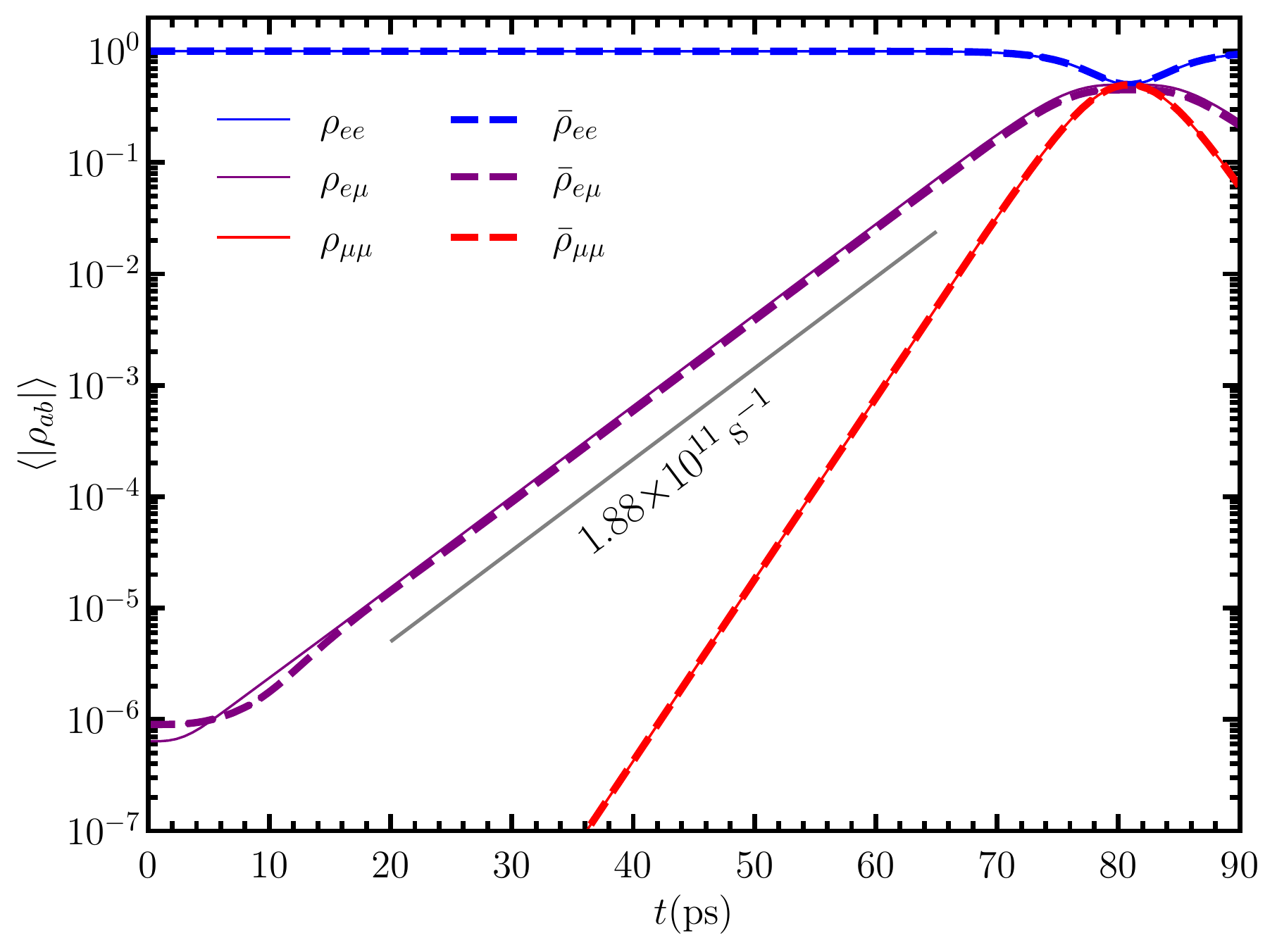}
    \caption{Inhomogeneous ($k \neq 0$) two-flavor two-beam fast flavor instability prepared in an initially nearly pure electron flavor state. The instability grows exponentially, as electron (blue) neutrinos (solid) and antineutrinos (dashed) rotate toward the pure muon state (red) and a quantum superposition of electron and muon flavor states (purple). The growth rate of the flavor off-diagonal component matches the prediction from linear stability theory well (gray line segment).}
    \label{fig:k_nonzero_fast_flavor_test}
\end{figure}

Finally, we perform a simulation of the inhomogeneous two-beam two-flavor fast flavor instability to test all of the previous components of the code together with the spatial advection of neutrinos. We choose the same neutrino masses and mixing angle as in Section~\ref{sec:homogeneous_two_beam_ffi}. In order to get the wavelength of the fastest growing mode to match the size of our box (i.e. $k=2\pi/(1\,\mathrm{cm})$), we set each of the neutrino and antineutrino number densities to 
\begin{equation}    
n=\frac{(m_2^2-m_1^2)c^4 /(2h\nu) + \hbar c k}{2\sqrt{2} G_F}=4.89\times10^{32}\,\mathrm{cm}^{-3} 
\end{equation}
such that the fastest growing mode has a wavelength of $1\,\mathrm{cm}$ for a neutrino energy of $50\,\mathrm{MeV}$ (see \cite{chakraborty_self-induced_2016} Equation~2.7). To isolate the fastest growing mode, we also perturb the initial conditions using $\mathrm{Re}(f_{e\mu})=\mathrm{Re}(\bar{f}_{e\mu})=10^{-6}\sin(k z_0)$, where $z_0$ is the initial position of the particle. The results are shown in Figure~\ref{fig:k_nonzero_fast_flavor_test}. As for the homogenous case in Section~\ref{sec:homogeneous_two_beam_ffi}, we see neutrinos (solid) and antineutrinos (dashed) mix from the nearly pure initial electron flavor state (blue) into the muon flavor state (red) and a quantum superposition of muon and electron flavor states (purple). We see the growth of the off-diagonal component is exponential, with a growth rate matching the prediction of linear stability theory of $\mathrm{Im}(\omega) = (m_2^2-m_1^2)c^4/(2\hbar h\nu) + c k=1.88\times 10^{11}\,\mathrm{s}^{-1}$ between $t=33\,\mathrm{ps}$ and $53\,\mathrm{ps}$ with a relative error of at most $4.3\times10^{-3}$.
 
\section{Convergence}
\label{sec:convergence}
The calculations we perform here are on a very small length and time scale (typically $64~\mathrm{cm}$ domain size) compared to a supernova or neutron star merger. Thus, the quantity we are primarily interested in is the net amount of electron neutrinos that remain after the instability saturates, averaged over the entire domain and over time. We compute the spatial-temporal average of the neutrino and antineutrino density matrices $\langle N_{ab}\rangle$ and $\langle \bar{N}_{ab} \rangle$ as
\begin{equation}
    \langle N_{ab} \rangle = \frac{ \sum_\mathrm{i=1}^{N_\mathrm{zones}} \sum_\mathrm{j=j_\mathrm{start}}^{N_\mathrm{snapshots}} n_{ab,ij}}{N_\mathrm{zones} (N_\mathrm{snapshots}-j_\mathrm{start}+1)}
\end{equation}
where $i$ labels a grid zone and $j$ labels the simulation snapshot. $j_\mathrm{start}$ is the first snapshot after $t=3\,\mathrm{ns}$. This allows us to average only over times unaffected by the initial transitory phase of the instability growing and saturating. After saturation, the system is chaotic and thus impossible to demonstrate exact convergence. However, since the quantity significant for core-collapse and merger simulations is precisely $\langle N_{ab}\rangle$, this is the quantity we ensure is converged.

There are three parameters dictating the fidelity of these simulations. The first is the domain size of the simulation. Larger domains can host modes with longer wavelengths, which even if they are not the fastest growing unstable modes, they may still be unstable or interact with shorter-wavelength modes after saturation. The second is the number of grid cells. The background grid discretizes the neutrino potentials, so finer grid cells are required to allow modes with shorter wavelengths to be represented in the background potential. Finally, there is the number of computational particles per cell. More particles means that a larger number of unique particle directions are represented on the grid and modes with finer angular structure can be expressed by the simulation. We use only one particle per direction per cell, since we have found that placing multiple particles at different locations in each cell but moving in the same direction does not result in a solution significantly different from the one calculated using only one particle per direction per cell. In the simulations discussed below, we initialize the neutrino distribution as described in Section~\ref{sec:initialConditions} to assess what simulation parameters lead to a numerically converged result.

\begin{figure*}
    \centering
    \includegraphics[width=\linewidth]{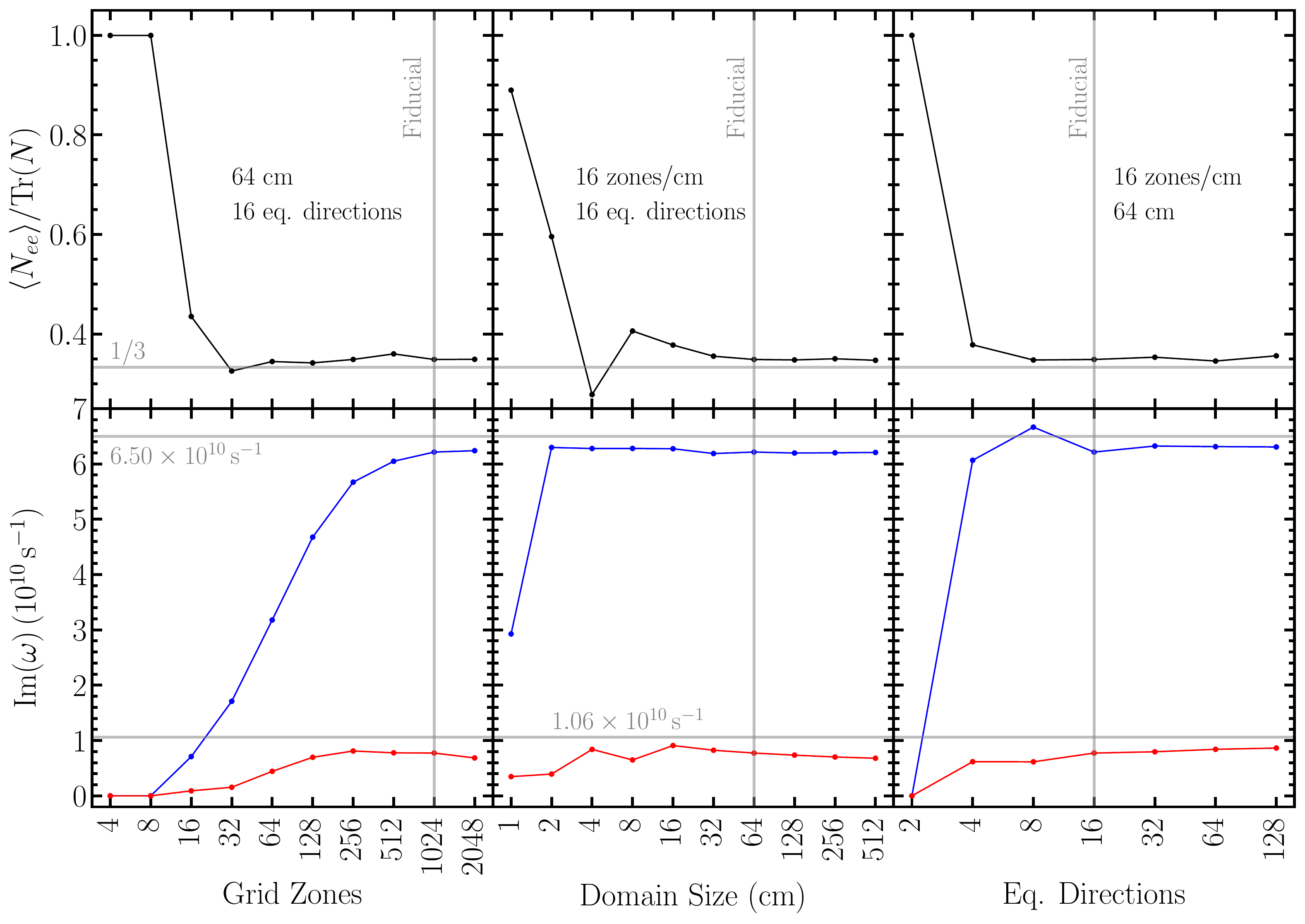}
    \caption{Convergence of the post-saturation electron neutrino flavor transformation (top) and fitted growth rate of the axially symmetric (bottom, blue) and asymmetric (bottom, red) modes in the fiducial simulation setup. The left column shows a varying number of grid cells covering a fixed domain of $64\,\mathrm{cm}$, the center column shows a varying domain size with a fixed number of grid cells per centimeter, and the right column shows a varying number of particles per cell. This number of directions is expressed in terms of the number of particle directions on the equator (in the $x-y$ plane). The total number of particles per cell is approximately $C^2/\pi$, where $C$ is the number of equatorial directions. 16 cells/cm, a domain size of 64 cm (1024 total cells), and 16 equatorial directions (92 particles per cell) are sufficient for convergence of the averaged post-saturation electron flavor content.}
    \label{fig:spatial_convergence}
\end{figure*}

The top left panel of Figure~\ref{fig:spatial_convergence} shows the space- and time-averaged fraction of $\nu_e$ in the post-saturation state under variations in the number of grid zones for a fixed domain size of 64 cm. Using eight or fewer grid zones causes the instability to never develop, leaving the final fraction of electron neutrinos equal to the initial conditions. As the number of grid zones increases beyond that, the saturated state appears to move increasingly toward an evenly mixed flavor distribution until 32 grid zones. Then the separation between the flavors increases slightly until 1024 grid zones, beyond which the results are mostly constant. Note that the numbers shown here are the values averaged over a simulation with only a duration of $5\times10^{-9}\,\mathrm{s}$, so this gap between electron and heavy lepton neutrinos is at least in part a result of not allowing the simulations to extend until momentum-space decoherence is complete. In the bottom left panel, we plot the fitted growth rate of the axially symmetric (blue) and asymmetric (red) modes along with the corresponding theoretical results from a linear dispersion analysis (gray horizontal lines). The growth rate of the symmetric mode appears to asymptote to the theoretical value. The growth rate of the asymmetric mode approaches the theoretical value until $n_z=256$ and then appears to slowly decrease. We expect that this is due in part to contamination from the underlying random perturbations, since the amplitude of the asymmetric mode is never large, and in part due to the limited resolution. The asymmetric mode does not significantly contribute to the post-saturation state, so we leave a more finely resolved simulation of the asymmetric mode to future work.

A similar progression occurs for changes in the domain size while keeping the size of each grid cell constant (middle panels in Figure~\ref{fig:spatial_convergence}). If the domain has an extent of only 1cm (4 grid cells) only a weak instability occurs, since the grid cannot contain the fastest growing mode. The amount of transformation over the course of the simulation varies up until a domain size of $64\,\mathrm{cm}$, after which it remains roughly constant. The growth rate of the symmetric mode is nearly correct at a $2\,\mathrm{cm}$ domain size, since it can nearly fit the fastest growing mode with a wavelength of $2.2\,\mathrm{cm}$. There are once again small variations until a domain size of $64\,\mathrm{cm}$, after which the growth rate remains relatively constant. The growth rate of the asymmetric mode once again proves to be difficult, peaking at a domain size of $16\,\mathrm{cm}$ and then slowly decreasing with increasing domain size. We expect this to be the result of the same reasons as for the behavior in the $n_z$ convergence test above.

Finally, we test the robustness of the results while varying the number of particles per cell, which also allows the particles to probe a larger number of directions. The leftmost point in the right panels of Figure~\ref{fig:spatial_convergence} represents a two-beam calculation. In each grid cell one particle is moving in the $+z$ direction and one particle is moving in the $-z$ direction. For the rest of the points, the horizontal axis describes the number of particle directions along the equatorial ($x-y$) plane. Particles are given directions according to Section~\ref{sec:initialConditions}. 4, 8, 16, 32, 64, and 128 equatorial directions correspond to 6, 24, 92, 378, 1056, and 6022 particles per cell, respectively. The amount of post-saturation flavor transformation (top panel) does not appear to be sensitive to the angular resolution of the simulation beyond 8 equatorial directions. The symmetric mode growth rate appears to vary little after 16 equatorial directions. Once again, the asymmetric mode growth rate is more difficult to resolve, but approaches the theoretical line with increasing angular resolution. It is interesting that such sparse sampling of the angular distribution can yield such accurate simulations of the FFI. This is in contrast to simulations of collective oscillations that require thousands of angular bins \cite{duan_self-induced_2011}.

\begin{figure}
    \centering
    \includegraphics[width=\linewidth]{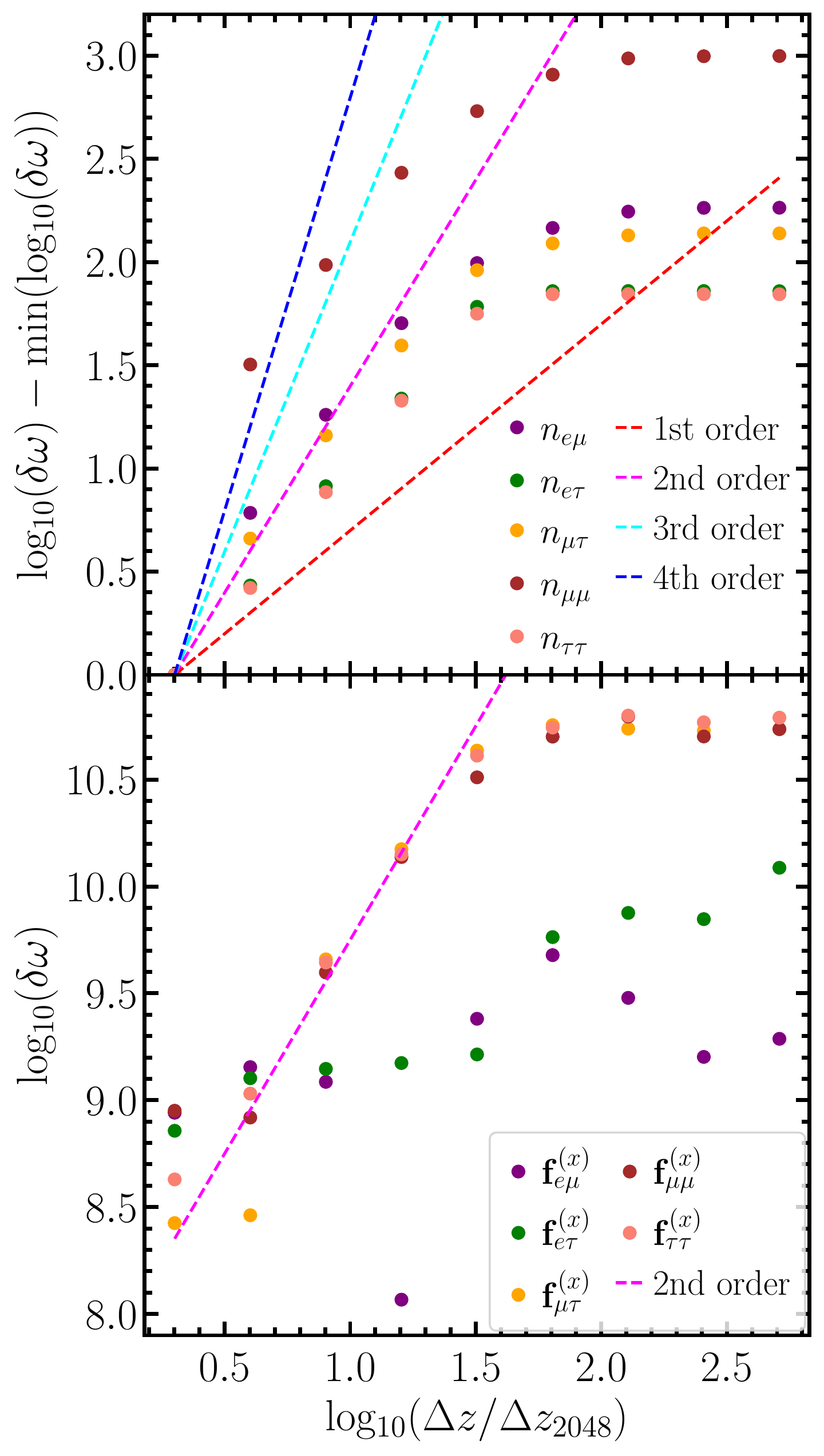}
    \caption{Convergence of the error in the growth rate of the neutrino number density (top panel) and $\hat{x}$ component of the flux (bottom panel) with increasing grid cell size $\Delta z$ relative to the cell size in the $n_z=2048$ simulation ($\Delta z_{2048}=0.03125\,\mathrm{cm}$). Here $\delta \omega = |\mathrm{Im}(\omega-\omega_{2048})|$, where $\omega_{2048}$ is the fitted growth rate in the $n_z=2048$ simulation. The color of each point denotes the component of the flavor matrix. For reference, scaling relations for convergence of order 1, 2, 3, and 4 are shown as dashed lines. {\tt Emu} achieves better than first order convergence for the growth rate of all distribution quantities except $\mathbf{f}_{e\mu}$ and $\mathbf{f}_{e\tau}$.}
    \label{fig:convergence_order}
\end{figure}

We also investigate our integration method's order of space-time convergence. We use the data set shown in the leftmost column of Figure~\ref{fig:spatial_convergence} where we kept the extent of the domain constant while varying the number of grid zones, thus varying the grid resolution $\Delta z$. We measure convergence by measuring the instability growth rate during the linear phase in the time interval $[0.15, 0.25]\,\mathrm{ns}$, firmly centered in the linear regime as shown in Figure~\ref{fig:1d_linear}. We take as a reference point for computing the convergence error the highest resolution simulation shown in Figure~\ref{fig:spatial_convergence} corresponding to $n_z = 2048$ grid cells in the domain. We denote the grid resolution of our $n_z = 2048$ simulation as $\Delta z_{2048}$ and each other grid resolution where we measured convergence error as $\Delta z$. We then denote the convergence error of the growth rates in the lower-resolution simulations as $\delta \omega = |\mathrm{Im}(\omega - \omega_{2048})|$ and compute this error for the same neutrino densities $n_{ab}$ and fluxes $\mathbf{f}_{ab}^{(x)}$ shown growing in Figure~\ref{fig:1d_linear}. In the top panel of Figure~\ref{fig:convergence_order}, we shift all logarithmic growth rate errors of each component of $n_{ab}$ by the minimum logarithmic error for that component across all resolutions. This permits us to clearly compare our achieved convergence order to representative lines (dashed) for first (red), second (magenta), third (cyan), and fourth (blue) order convergence rates. We see that towards high resolutions, density components $n_{e\mu}$ (purple), $n_{\mu\tau}$ (yellow), and $n_{\mu\mu}$ (brown) all converge to second order or better. Exceptions are $n_{e\tau}$ (green) and $n_{\tau\tau}$ (salmon), which converge between first and second order. We note that although we are using a fourth order Runge-Kutta method as our time integrator, the PIC deposition and interpolation shape functions we use reduce the achievable space-time convergence to second order, as we discuss in Section~\ref{sec:pic_deposition}. 

In the bottom panel of Figure~\ref{fig:convergence_order} we next show the unscaled convergence error $\delta \omega$ for the axially asymmetric flux $\mathbf{f}_{ab}^{(x)}$ with the same colors representing each component as in the top panel. Here we see that at higher resolutions, $\mathbf{f}_{\mu\tau}^{(x)}$, $\mathbf{f}_{\mu\mu}^{(x)}$, and $\mathbf{f}_{\tau\tau}^{(x)}$ converge to second order, although $\mathbf{f}_{e\mu}^{(x)}$ and $\mathbf{f}_{e\tau}^{(x)}$ do not. We note nevertheless that at the fiducial resolution of $n_z = 1024$ (the leftmost set of points in the lower panel of Figure~\ref{fig:convergence_order}), the size of the error $\delta \omega$ corresponds to $\approx 10\%$ error compared to the axially asymmetric growth rate controlling $\mathbf{f}_{e\mu}^{(x)}$ and $\mathbf{f}_{e\tau}^{(x)}$. The relative error for other components is lower, as other components grow at either the symmetric growth rate or twice that rate. Convergence studies that explore the joint grid resolution and angular resolution space may shed additional light on what determines the convergence requirements for $\mathbf{f}_{e\mu}^{(x)}$ and $\mathbf{f}_{e\tau}^{(x)}$, and we leave this for future work.

These results justify our choice of fiducial simulation parameters. However, it should still be noted that these simulations are too short to probe the long-term behavior of the post-saturation nonlinear evolution.

\end{document}